\documentclass[aip,rsi,reprint]{revtex4-1}
\pdfoutput=1
\usepackage{graphicx,bm,amsmath,amssymb}
\usepackage{dcolumn}
\usepackage{color}

\newcommand{\rn}{\mathbf{r}_{\nu}}
\newcommand{\rnp}{\mathbf{r}_{\nu +1}}
\newcommand{\kn}{\mathbf{k}_{\nu}}
\newcommand{\qn}{\mathbf{q}_{\nu}}
\newcommand{\en}{\mathbf{e}_{\nu}}
\begin{document}

\title{Spatially-resolved measurements of micro-deformations in
  granular materials using Diffusing Wave Spectroscopy}



\author{Axelle Amon}
\email[]{axelle.amon@univ-rennes1.fr}
\author{Alesya Mikhailovskaya}
\author{J\'er\^ome Crassous}
%\thanks{}
\affiliation{Universit\'e de Rennes 1, Institut de Physique de Rennes
  (UMR UR1-CNRS 6251), B\^{a}t.~11A, Campus de Beaulieu, F-35042
  Rennes, France}


\date{\today}

\begin{abstract}
This article is a tutorial on the practical implementation of a method
of measurement of minute deformations based on multiple
scattering. This technique has been recently developed and has proven
to give new insights on the spatial repartition of strain in a
granular material. We provide here the basics to understand the method
by giving a synthetic review on Diffusing Wave Spectroscopy and
multiple scattering in granular materials. We detail a simple
experiment using standard lab equipment to pedagogically demonstrate
the implementation of the method. Finally we give a few examples of
measurements that have been obtained in other works to discuss the
potential of the method.
\end{abstract}


\maketitle %\maketitle must follow title, authors, abstract and \pacs

%%%%%%%%%%%%%%%%%%%%%%%%%%%%%%%%%%%%%%%%%%%%%%%%%%%%%%%%%%%%%%%%%%%%%%%%%%%%%%%%%%%%%%%%%%%%%%%%%%%%%%%%%%%%%%%%%%%%%%%%%%%%%%%%%%%%%%%%%%%%%%%%%%%%%%%%%%%%%%%%%%%%%%%%%%%%%%%%%%%%%%%%%%%%%%%%%%%%%%
\section{Introduction}
Numerous types of disordered media, as foams, grains or concentrate
colloidals suspensions strongly scatter light because of their
heterogeneities. This feature is usually considered as an obstacle to
accurate optical diagnoses in those materials. However information can
be extracted from the light multiply scattered by a sample, giving
access to local minute relative
displacements~\cite{Weitz1993,Maret1997}. The possibility of such
measurements in strongly scattering media have raised a lot of
interest for the study of soft glassy materials as those systems are
often turbid. The earliest works devoted to Brownian motion in
colloidal suspensions~\cite{Maret1987,Pine1988} have been rapidly
extended to other scattering media as
foams~\cite{Durian1991,Hohler1997}, emulsions~\cite{Hebraud1997} or
granular materials~\cite{Menon1997}. For all those systems,
measurements based on multiple scattering have proven to be crucial
for the understanding of their dynamics~\cite{Hohler2014}. A potential
drawback of those methods is that intrinsically only mean quantities
can be obtained as the underlying multiple scattering process produce
a natural averaging of the information. Numerous works have been
devoted to improvements and variations of the original
method~\cite{Zakharov2009} in particular with the goal of
characterizing non-ergodic and heterogeneous
dynamics~\cite{Cipelletti2003,Mayer2004,Duri2009,Dixon2003,Bandyopadhyay2005,Erpelding2008}. Of
particular interest here are the methods which allow to achieve
spatial
information~\cite{Erpelding2008,Zakharov2010,Sessoms2010a}. Full-field
methods of measurement of strain and displacement based on speckle
interferometry has been developped since the
70's~\cite{Dainty1984,Rastogi2000}. Those methods are based on the
measurement of planar displacements and deformation of a rough
surface. In the case of the method presented here, the light
penetrates inside the material and the information collected is not
merely surfacic~\cite{Erpelding2013}.

In the present article we show in details the implementation of a
method of measurement of tiny deformation based on multiple
scattering. The method is based on the measurement of auto-correlation
function of the light intensity multiply scattered by a strongly scattering
material. We provide here the basics needed to understand the method
and implement it practically in a lab.

Part II is an introduction to Diffusing Wave Spectroscopy (DWS) giving
the essentials to understand the method. Part III covers the specific
case of granular materials. Scattering processes in a granular
assembly is discussed. The effect of correlated and uncorrelated
motion of the scatterers on the phase shift is presented, leading to a
general expression of the auto-correlation function of the scattered
intensity as a function of the strain field. The principle of the
method for obtaining a spatially resolved strain maps is then
presented. Part IV is a tutorial describing in full details the
implementation of the method in a simple experiment based on heat
conduction in a granular material. It provides all practical details
of the adjustments that have to be done to perform the experiment
properly. The last part presents a short review of the literature of
measurements that have been done on granular systems using the method
and indicating its potential and limitations.

\section{DWS theory}
\subsection{General principle} \label{sec:DWS_princ}

\begin{figure}[hbtp]
\begin{center}
\includegraphics[width=.8\columnwidth]{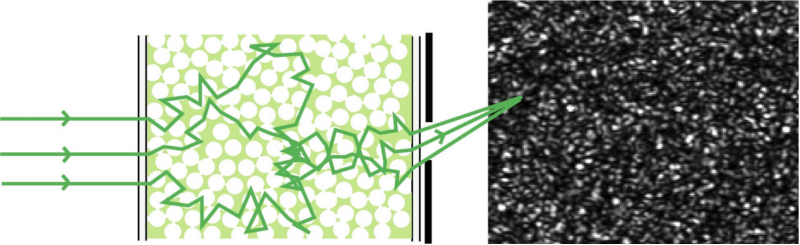}
\caption{Left: scattering in a packing of glass beads. Light rays
  inside the material follow complicated paths made of numerous
  reflection and refraction events. Right: the collected rays interfere
  resulting in a speckle image.}
\label{light_speckle}
\end{center}
\end{figure}

In a strongly scattering material, light rays follow different paths,
each path being composed of numerous scattering events (see
Fig.~\ref{light_speckle}). When the source is coherent, the light
transmitted through or backscattered from a disorder system gives rise
to constructive and destructive wave interference and the collected intensities display a speckle
figure (see Fig.~\ref{light_speckle}). The scattered light rays have
performed a random walk inside the material and thus have explored a
part of the bulk of the material.

If the system has an internal dynamics (e.g., Brownian motion for a
colloidal suspension) the speckle figure will change with time. The
principle of the \emph{Diffusing-Wave Spectroscopy} (DWS) is to
analyze the fluctuations of the scattered intensity in order to
extract information about the structure or dynamics of the
system~\cite{Pine2000,Weitz1993}.

The principle of the analysis is based on the calculation of the
auto-correlation function $g_I$ of the scattered intensities. The
correlations are calculated between two states of the sample that we
will call 1 and 2 in the following:
\begin{equation}
g_I (1,2) =\frac{\left\langle I_1 I_2 \right\rangle}{\left\langle
I_1 \right\rangle \left\langle I_2 \right\rangle} \label{eq:def_gI}
\end{equation}
The average operation $\langle \cdot \rangle$ in Eq.~\eqref{eq:def_gI}
can be performed over time (it is then supposed that the phenomenon
studied is ergodic), or over an ensemble of
speckles~\cite{Scheffold2007}. In multi-speckles
method~\cite{Viasnoff2002}, light intensity is collected on a CCD
camera, each speckle being an independent representation of the same
random process and an ensemble average is obtained by averaging over
the pixels of the camera.
\begin{figure}[hbtp]
\begin{center}
\includegraphics[width=1.\columnwidth]{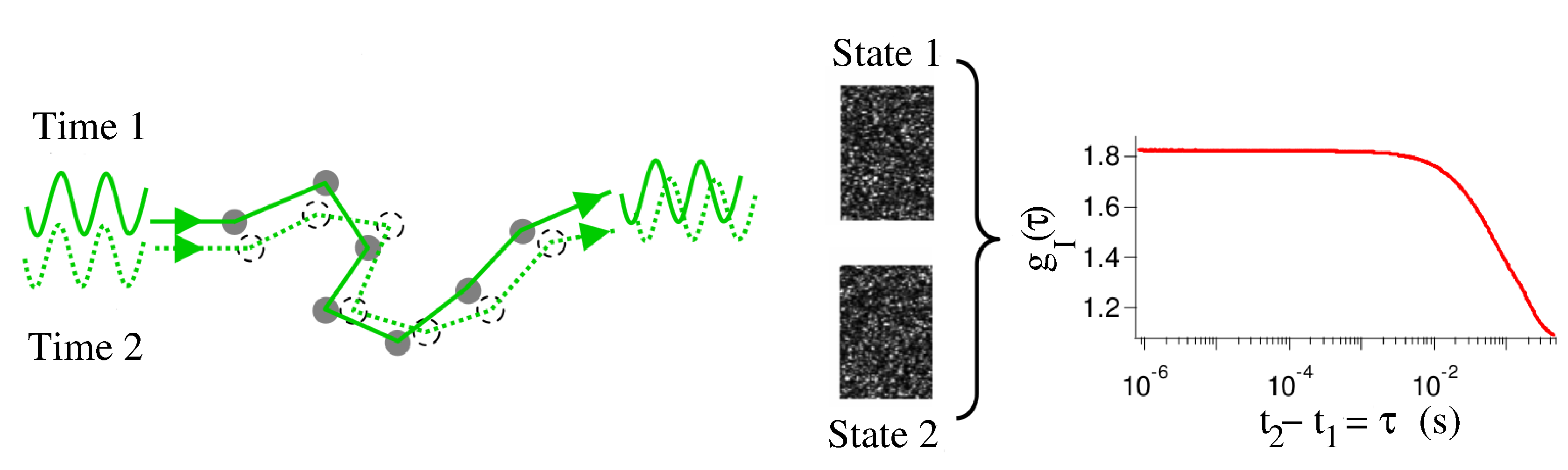}
\caption{From~\cite{Erpelding2010b}: principle of DWS. Because of some
  inner dynamics, a relative modification of the positions of the
  scatterers occurs during time. Each configuration of the structure
  of the material gives a different speckle image. The calculated
  correlation function of the transmitted intensity $g_I(\tau)$
  between the intensities $I_1=I(t)$ and $I_2 = I(t+\tau)$ decreases
  with the time lag $\tau$. The curve has been obtained during the
  aging of shaving foam.}
\label{def_GI}
\end{center}
\end{figure}

The loss of correlation measured between two speckle figures
corresponds to two different states generated by the inner dynamics in
the material (see Fig.~\ref{def_GI}).  When single scattering is at
play, information can be directly extracted when collecting the light
at a given angle from the deviation of the wavevector. In the case of
multiple scattering the wavevectors have experienced numerous
deviations so that the dependence of the signal on the scattering
angle is lost and no specific information can be inferred. Still, the
modeling of the light propagating in the material as a diffusion
process~\cite{Ishimaru1978} makes it possible to characterize the
dynamics in the system.  Although the diffusion process causes the
loss of most of the detailed information about the material in which
it propagates, this process is also at the origin of the unmatched
sensitivity of the method. Usually, the sensitivity of an
interferometric method is of the order of the wavelength of the
coherent source used. Indeed, a loss of correlation between two
interferometric figures corresponds typically to the change from
constructive interference to destructive one, \emph{i.e.}  to a change
between the ray paths of the order of one wavelength. But as multiple
scattering implies a large number of scattering events, decorrelation
will occur when the scatterers will have move only of a fraction of
the wavelength. Consequently, relative displacements of a few
nanometers of the scatterers are measurable~\cite{Weitz1993}.

\subsection{Amplitude and intensity correlation
  functions} \label{sec:cor_func}
Considering the field at a point on a sensor, its amplitude $E$ is the
sum of a large number of rays $E(t) = \sum_{\alpha}
\mathcal{E}_{\alpha}(t)$. The correlation function of the amplitudes
is:
\begin{equation}
g_E(1,2)=\frac{\left\langle E_1 E_2^*\right\rangle}{\left\langle
|E_1|\right\rangle \left\langle |E_2|\right\rangle},
\end{equation}
which implies the cross term:
\begin{equation*}
E(t_1) E^*(t_2) = \sum_{\alpha} \mathcal{E}_{\alpha}(t_1)
\mathcal{E}^*_{\alpha}(t_2) + \sum_{\alpha \neq \beta}
\mathcal{E}_{\alpha}(t_1) \mathcal{E}^*_{\beta}(t_2).
\end{equation*}
When averaging, the contribution from $\sum_{\alpha \neq \beta}
\left\langle \mathcal{E}_{\alpha}(t_1) \mathcal{E}^*_{\beta}(t_2)
\right\rangle$ will vanish because the fields $\mathcal{E}$
originating from different paths ($\alpha \neq \beta$) can be
considered as uncorrelated. Consequently:
\begin{equation*}
\left\langle E(t_1) E^*(t_2) \right\rangle =  \sum_{\alpha}
\left\langle \mathcal{E}_{\alpha}(t_1) \mathcal{E}^*_{\alpha}(t_2)
\right\rangle \propto \sum_{\alpha} \left\langle e^{j \Delta \phi_\alpha} \right\rangle
\end{equation*}
with $\Delta \phi_\alpha = \phi_\alpha(t_2) - \phi_\alpha(t_1)$. This
result means that the statistical properties of the fluctuations
depend only on the phase variation of each paths. Consequently,
understanding the form of the correlation function necessitates to
calculate the typical phase variation of a path.

Experimentally the correlation is calculated over the
intensities. When the scattered field $E$ has a Gaussian distribution,
the correlation functions on amplitudes, $g_E$, and the ones on
intensities, $g_I$, are linked by the Siegert
relation~\cite{Berne2000}:
\begin{equation}
g_I(1,2)=1+\beta_e |g_E(1,2)|^2 \label{eq:siegert}
\end{equation}
where $\beta_e$ is an experimental constant of order unity depending
on the details of the experimental setup~\cite{Berne2000}.

The function $g_E(1,2)$ depends on the phase variation on each of path
between the states 1 and 2. A variation of length $\Delta s$ of a path
of length $s$ leads to a phase variation $\Delta \phi_s = k \Delta s$
for the light ray following this path, with $k=2\pi/\lambda$ the
wavevector of the light, and $\lambda$ its wavelength. In the multiple
scattering limit, each path is composed of numerous scattering events
and the correlation function of the scattered field can be expressed
as~\cite{Weitz1993,Pine1990}
\begin{equation}
g_E(1,2)=\int_s \mathcal{P}(s) \left\langle
e^{j \Delta \phi_s(1,2)}\right\rangle ds \label{eq_int_gE}
\end{equation}
where $\mathcal{P}(s)$ is the probability for an optical path to have
the length $s$. This distribution can be calculated from the diffusive
equation knowing the boundary conditions of the
experiment~\cite{Ishimaru1978}. The quantity $\left\langle e^{j \Delta
  \phi_s(1,2)}\right\rangle$ is the contribution of a path of length
$s$ to the variation of the electric field between the states $1$ and
$2$ and the average $\left\langle \cdot \right\rangle$ is done over
all the paths of length $s$.

All the information about the deformation or dynamics of the material
is contained in the term $\left\langle e^{j \Delta
  \phi_s(1,2)}\right\rangle$~\cite{Maret1987}. The number of
scattering events in each paths is large so that by the central limit
theorem, $\Delta \phi_s$ is a random variable, and:
\begin{equation}
\left\langle e^{j \Delta \phi_s}\right\rangle = e^{j \langle \Delta
  \phi_s \rangle} \times e^{- \frac{\langle \Delta \phi_s^2 \rangle -
    \langle \Delta \phi_s \rangle^2}{2}}. \label{eq:exp_Dphi}
\end{equation}
In the next section, the phase shift obtained for a deformed material
will be detailed.

%%%%%%%%%%%%%%%%%%%%%%%%%%%%%%%%%%%%%%%%%%%%%%%%%%%%%%%%%%%%%%%%%%%%%%%%%%%%%%%%%%%%%%%%%%%%%%%%%%%%%%%%%%%%%%%%%%%%%%%%%%%%%%%%%%%%%%%%%%%%%%%%%%%%%%%%%%%%%%%%%%%%%%%%%%%%%%%%%%%%%%%%%%%%%%%%%%%%%%
\section{Imaging granular materials}

%%%%%%%%%%%%%%%%%%%%%%%%%%%%%%%%%%%%%%%%%%%%%%%%%%%%%%%%%%%%%%%%%%%%%%%%%%%%%%%%%%%%%%%%%%%%%%%%%
%%%%%%%%%%%%%%%%%%%%%%%%%%%%%%%%%%%%%%%%%%%%%%%%%%%%%%%%%%%%%%%%%%%%%%%%%%%%%%%%%%%%%%%%%%%%%%%%
\subsection{How scattering works in a granular material}

\begin{figure}[hbtp]
\begin{center}
\includegraphics[width=\linewidth]{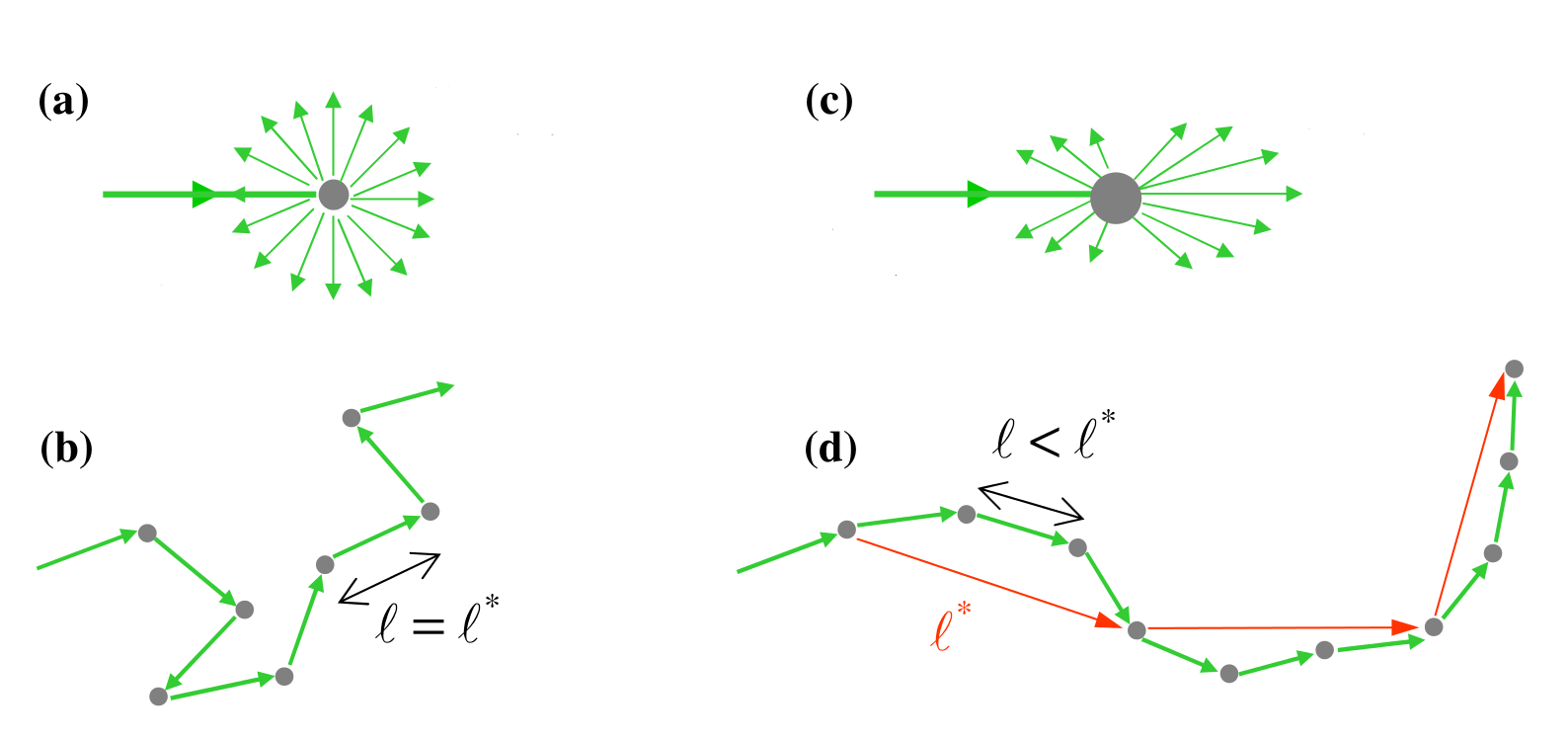}
\caption{From~\cite{Erpelding2010b}. (a) Isotropic scattering. (b)
  Light propagation can be modeled as a random walk of step
  $\ell=\ell^*$ with no correlation of orientation. (c) Anisotropic
  scattering. Because of the preferential direction of scattering, the
  random process has a persistent length. The total loss of memory is
  done on a length $\ell^*$ which counted several scattering steps
  $\ell$.}
\label{fig:l_star}
\end{center}
\end{figure}

The measured correlation functions of electromagnetic fields are
related to the phase fluctuations according to eq.~\eqref{eq_int_gE}
in which the path length distribution $\mathcal{P}(s)$ needs to be
known. This distribution is related to the transport of light into the
granular material. Multiple light scattering during its propagation
into the medium is characterized by two different lengths. First, the
average distance $\ell$ traveled by a photon in between successive
scattering events. In the limit of a very high multiple scattering,
photons are scattered a large number of times, and paths are random
walks. Another length is then introduces to describe this random walk:
the transport mean free path of the light in the material,
$\ell^*$. The transport light $\ell^*$ is different from the mean free
path $\ell$ in the case when light is not scattered isotropically by
the scatterers (see Fig.~\ref{fig:l_star}). A typical example of such
anisotropic scattering is the Mie scattering, i.e. the scattering by a
dielectric sphere of radius of the order or larger than the
wavelength~\cite{Hulst1981}. The light is then scattered
preferentially in the forward direction so that several scattering
events are necessary for the photon to lose the memory of its initial
orientation.
The relationship between $\ell$ and $\ell^*$ when scattering
  events are independent is:
\begin{equation}
\ell^* = \frac{\ell}{\langle 1 - \cos \theta \rangle}\label{eq:l_star}
\end{equation}
where $\theta$ is the scattering angle. For dense suspensions, this
expression has to be modified to take into account the structure
factor~\cite{Wolf1988}. Extension of~\eqref{eq:l_star} to packing of
particles that are large compared to the optical wavelength is
unclear. The use of the Mie scattering theory for estimating $\ell$
and $\langle \cos \theta \rangle$ is probably only approximative
because the scattered cannot be considered in the far field limit (see
chapter 12 of~\cite{Hulst1981}).

To describe the optical properties of granular material, one of the
simplest model is an assembly of identical spherical dielectric beads
of radius $R$. There is no analytical solutions for the propagation of
electromagnetic waves through such medium. A simpler approach is to
use geometrical optics to calculate light propagation through such
material. This is based on the fact that for granular material, the
beads diameter is large compared to the optical wavelength. In this
case, if the difference $\Delta n=n_{int}-n_{ext}$ of refractive
indices between the beads $n_{int}$ and surrounding media $n_{ext}$ is
such that $\Delta n R \gg \lambda$, geometrical optics should
describes light transport into beads assembly. For geometrical optics,
the only length scale is $R$, we then expect that
$\ell^*/R=f(n_{int},n_{ext},\varphi)$ where
$f(n_{int},n_{ext},\varphi)$ is a non-dimensional function of the
refractive indices and of the solid volume fraction $\varphi$. The
function $f$ has been calculated analytically for a model of non
polarized rays into a disordered packing of
disks~\cite{Sadjadi2008}. This function has been also determined
numerically using a ray-tracing algorithm for a 3D packing of spheres
at $\varphi=0.64$~\cite{Crassous2007}. It is found that
$\ell^*/R\simeq 6.6$ for glass beads ($n_{int}=1.51$) dispersed in air
($n_{ext}=1$).

\begin{table*}[htbp] \centering
\begin{tabular}{|p{2.5cm}|p{2.5cm}|p{2.5cm}|p{2.cm}|p{2.5cm}|}
\hline
Authors & $d$~($\mu m$) & $\ell^*/d$ & $\ell_a$~($mm$) \\
\hline
Menon~{\it et al.}\cite{Menon1997} &  $95$ & 7.5 & 12 \\
\hline
Lemieux~{\it et al.}\cite{Lemieux2000} & $330$ & 10 & \\
\hline
Dixon~{\it et al.}\cite{Dixon2003} & $780$ & 4 & \\
\hline
Djaoui~{\it et al.}\cite{Djaoui2005} & $51$ & 12.9 & \\
\hline
Crassous~{\it et al.}\cite{Crassous2007} & $67$ & 2.8 & 27\\
\hline
Crassous\cite{Crassous2009} & $80$ & 4.1 & 6 \\
\hline
\end{tabular} \caption{Reported values for $\ell^*$ and $\ell_a$ for
  glass beads of diameter $d$ in air.} \label{table1}
\end{table*}

In multiple scattering regime, light propagation can be described
using the diffusion approximation. Solving this equation permits to
find the function $P(s)$ for various geometries~\cite{Weitz1993}. The
energy density of light $\mathcal{U}_e$ then verifies a diffusion
equation
\begin{equation}
{\partial \mathcal{U}_e \over \partial t}=D \nabla^2
\mathcal{U}_e\label{eq:diff}
\end{equation}
where the the diffusion coefficient may be expressed as $D=v\ell^*/3$,
with $v$ the light velocity into the medium. In addition to $\ell^*$,
three other lengths may be defined~\cite{Weitz1993}. First, the
absorption length of photons into the material $\ell_a$ might needed
to be taken into account. Second, the initial condition of the
diffusion equation is approximative as the source becomes diffusive
only after a few steps inside the material. This leads to define a
penetration length, $\ell_0$, which is the distance into the sample
where the source must be located for taking into account this
randomization. Finally, the boundary conditions necessary to solve the
diffusion equation are approximative~\cite{Ishimaru1978} so that one
needs to define the distance outside the sample where the density of
light $\mathcal{U}_e$ extrapolates to zero. This distance is the
extrapolation length $\ell_e$.

Measuring $\ell^*$ in a granular packing is hampered by the absence of
Brownian motion: it can not be extracted from the decay time of the
temporal auto-correlation function, and other methods can be difficult
to implement as detailed in~\cite{Leutz1996}. Usually, the measurement
is done using transmission techniques~\cite{Leutz1996}. There is thus
few reported values of $\ell^*$ from experiments. Table~\ref{table1}
gathers a compilation of the experimental values of $\ell^*/d$ for
glass beads of diameter $d$ dispersed in air. Depending on the authors
and on the methods, this ratio shows significant variations,
underlining the difficulty to measure $\ell^*$.  To our knowledge
there is no reported experiments where $\ell_0$ and $\ell_e$ have been
measured for glass beads. At first approximation, the values
$\ell_0=\ell^*$ and $\ell_e=2 \ell^*/3$ may be taken.  The
experimental values of the absorption length $\ell_a$ are reported in
Table~\ref{table1}.

%%%%%%%%%%%%%%%%%%%%%%%%%%%%%%%%%%%%%%%%%%%%%%%%%%%%%%%%%%%%%%%%%%%%%%%%%%%%%%%%%%%%%%%%%%%%%%%%%
\subsection{Displacement fields and phase variation}

In this part we discuss the computation of the phase shift depending
on the motion of the scatterers. We have seen in
Sec.~\ref{sec:cor_func} (see Eq.~\eqref{eq_int_gE}) that by computing
the phase shift $\phi_p (t_2) - \phi_p (t_1)$ of a path $p$ composed
of $N_p$ scattering events between the states obtained at times $t_1$
and $t_2$, one can deduce the expression of $g_I(t_1,t_2)$.

First we will consider the case of an affine deformation of the
scattering material, then we will move to the case of uncorrelated
motion of the scatterers and finally we will discuss the superposition
of those two kinds of motion.

\begin{figure}[htbp]
\includegraphics[width=1.\columnwidth]{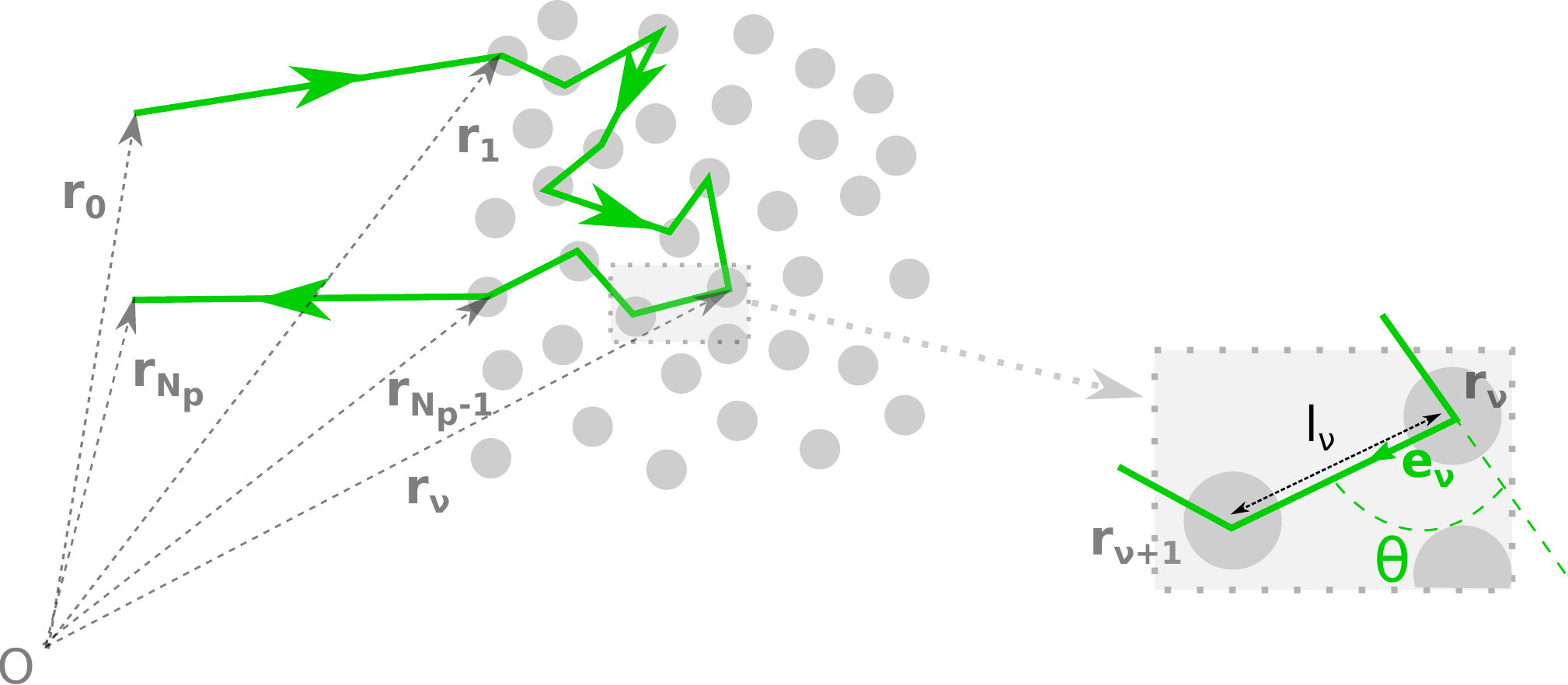}%
\caption{A light ray path $p$ composed of
  $N_p$ scattering events.\label{fig:def_r}}%
\end{figure}

All along this part we will use the following notations (see
Fig.~\ref{fig:def_r}): $\rn$ is the position of $\nu^{th}$ scatterer
and $\kn$ is the wavevector of the light after the $\nu^{th}$
scattering event, $\kn = k \en$, where $\en = \frac{\rnp -
  \rn}{l_{\nu}}$ and $l_{\nu} = \| \rnp - \rn\|$. The total absolute
phase along a path is then:
$$ \phi_p = \sum_{\nu =0}^{N_p} \kn \cdot \left(
\rnp - \rn \right) = k \sum_{\nu =0}^{N_p}
l_{\nu}.$$

%%%%%%%%%%%%%%
\subsubsection{Affine deformation field}
In the case of an affine deformation taking place between times $t_1$
and $t_2$, the displacement of the scatters is described by a
displacement field between two states $\bm{u} = \rn(t_2) -
\rn(t_1)$. From this displacement field a strain tensor can be
defined: $U_{ij} = \frac{1}{2} \left( \frac{\partial u_i}{\partial
  x_j} + \frac{\partial u_j}{\partial x_i} \right)$.

\begin{figure}[htbp]
\includegraphics[width=0.6\columnwidth]{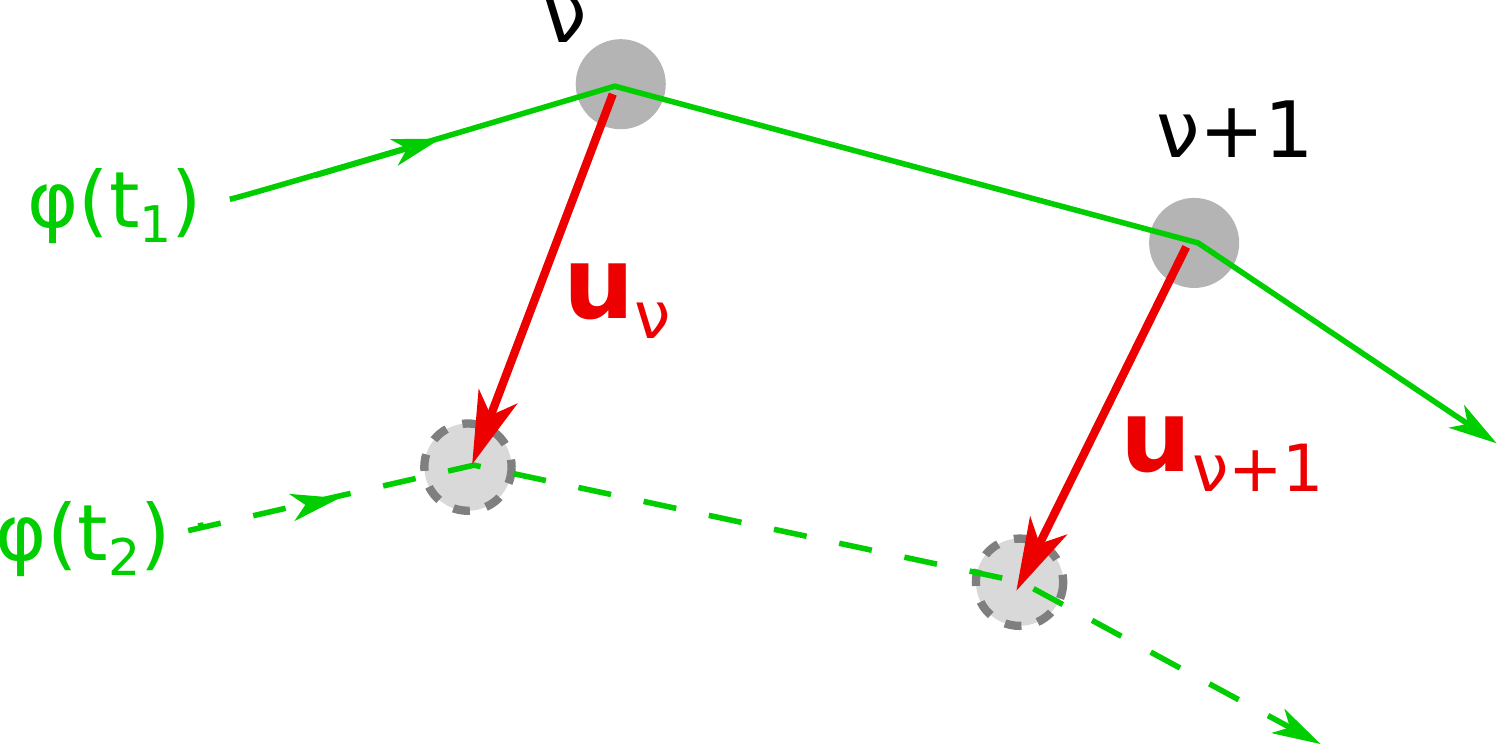}%
\caption{Schematic of the phase variation on a section of a path in
  the case of a displacement field $\bm{u}$.\label{fig:detail_phase_A}}%
\end{figure}
The total phase variation $\Delta \phi_A = \phi (t_2) - \phi (t_1)$
due to the affine motion between two states can be
computed~\cite{Bicout1991,Bicout1993,Bicout1994}:
\begin{eqnarray*}
  \Delta \phi_A &=& \sum_{\nu =0}^{N_p} \kn \cdot \left[
    \left(\rnp(t_2) - \rnp(t_1) \right) - \left( \rn(t_2) - \rn(t_1)
    \right) \right]\\
  &=&\sum_{\nu =0}^{N_p} \kn \cdot \left[ \bm{u}(\rnp) -
    \bm{u}(\rn)\right]
\end{eqnarray*}
where the variation of $\kn$ between the two states has been neglected
as it gives rise to second order terms~\cite{Weitz1993}. If $\bm{u}$
varies slowly over the length scale $\langle l_{\nu} \rangle = \ell$,
we have:
$$ \bm{u}(\rnp) = \bm{u}(\rn + l_{\nu} \en) \approx \bm{u}(\rn) +
l_{\nu} (\en \cdot \nabla) \bm{u}(\rn)$$ so that
$$\bm{u}(\bm{r}_{\nu + 1}) - \bm{u}(\bm{r}_{\nu}) \approx l_{\nu}
\sum_{i,j} e_{\nu,i} e_{\nu,j} U_{ij} (\bm{r}_{\nu}).$$ In the
multiple scattering limit, using the fact that the orientations of
wave vectors are not correlated with the direction of the strain
tensor, we
have~\cite{Bicout1991,Bicout1993,Bicout1994,Djaoui2005,Crassous2007}:
\begin{eqnarray*}
\langle \Delta \phi_A \rangle &=& k N_p \langle l_\nu \rangle
\sum_{i,j} \langle e_{{\nu},i} e_{{\nu},j} \rangle \langle U_{ij}\rangle\\
\langle \Delta \phi_A^2 \rangle &=& k^2 \sum_{\nu,\nu'} \langle l_\nu
l_{\nu'} \rangle \sum_{ij,i'j'} \langle e_{{\nu},i} e_{{\nu},j}
e_{{\nu'},i'} e_{{\nu'},j'} \rangle \langle U_{ij} U_{i'j'} \rangle
\end{eqnarray*}
Because of the isotropic orientation of the different directions of
scattering, $\langle \Delta \phi \rangle$ and $\langle \Delta \phi^2
\rangle$ depend only on the isotropic invariants of the strain
tensor. Moreover, in the multiple scattering limit, we expect that the
variance of the phase shift varies linearly with the number of
scattering events and thus with the path length $s = N_p \langle
l_{\nu} \rangle$. Consequently, we expect:
\begin{eqnarray}
\langle \Delta \phi_A \rangle &=& \frac{1}{3} k s \text{Tr} \left(
\mathbf{U} \right)\\ \langle \Delta \phi_A^2 \rangle - \langle \Delta
\phi_A \rangle^2 &=& k^2 s \left[ (\beta - \chi) \text{Tr}^2 \left(
  \mathbf{U}\right) + 2 \beta \text{Tr} \left( \mathbf{U}^2\right)\right]
\end{eqnarray}
where $\beta$ and $\chi$ are lengths that scale as the bead radius and
which can be computed for a known scattering
process~\cite{Bicout1993,Bicout1994}. A numerical evaluation of those
lengths using ray tracing has been done by
Crassous~\cite{Crassous2007}. He has shown that, for not to large
contrast of indices between the beads and the surrounding medium, the
values obtained are in agreement with the expression for Mie
scatterers with no correlation of distance between successive
scattering events obtained by Bicout \emph{et
  al.}~\cite{Bicout1993,Bicout1994}. We thus expect values close to
$\beta = \frac{2l^*}{15}$ and $\chi = 0$.

%%%%%%%%%%%%%%%
\subsubsection{Uncorrelated motion of the diffusers}
In the case when the displacement of the scatterers is purely random,
we have
$$\rn(t_2) = \rn(t_1) + \bm{\xi}(\rn),$$
with the components $\xi_i$ of the vector $\bm{\xi}$ random variables
of zero mean, $\langle \xi_i \rangle =0$.
\begin{figure}[htbp]
\includegraphics[width=0.7 \columnwidth]{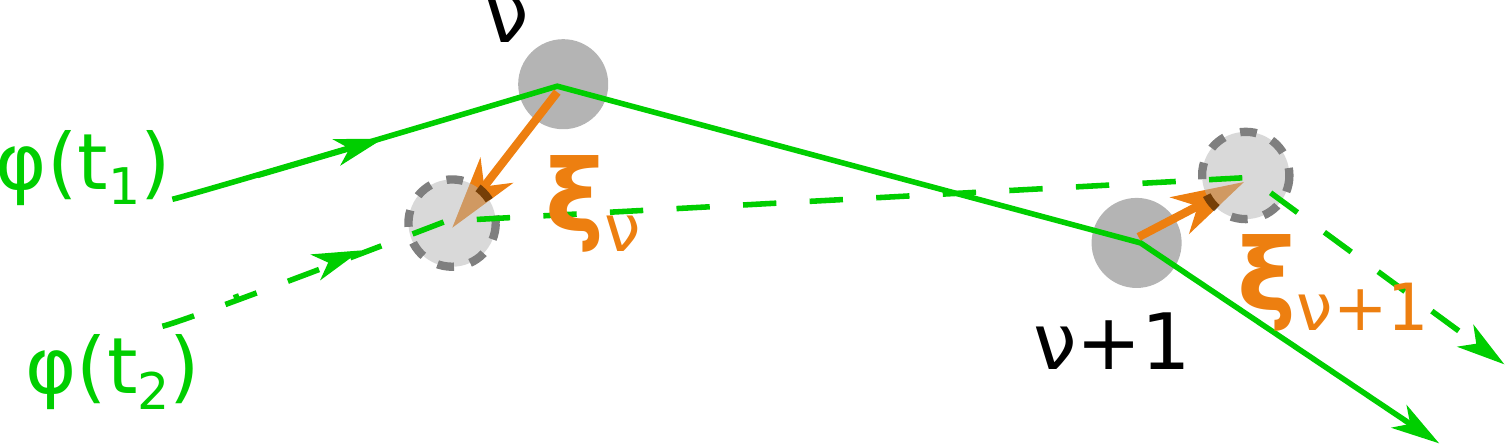}%
\caption{Schematic of the phase variation on a section of a path in
  the case of uncorrelated motion of the scatterers\label{fig:detail_phase_NA}}%
\end{figure}
The phase variation due to the uncorrelated motion of the scatterers
$\Delta \phi_{NA}$ can then be computed~\cite{Weitz1993}:
\begin{eqnarray*}
  \Delta \phi_{NA}&=& \sum_{\nu =0}^{N_p} \kn \cdot \left[
    \left(\rnp(t_2) - \rnp(t_1) \right) - \left( \rn(t_2) - \rn(t_1)
    \right) \right]\\
&=& \sum_{\nu =1}^{N_p} \left[\bm{k}_{\nu + 1}  - \kn \right]\cdot
  \bm{\xi}(\rn) = \sum_{\nu =1}^{N_p} \qn \cdot
  \bm{\xi}(\rn)
\end{eqnarray*}
where $\bm{q}_{\nu} = \bm{k}_{\nu} - \bm{k}_{\nu-1}$ is the scattering
vector.

As the scattering vectors $\qn$ and the random motions $\bm{\xi}(\rn)$ are
independent, we have
$$ \langle \bm{q}_{\nu} \cdot \bm{\xi}(\mathbf{r}_{\nu}) \rangle =
\sum_i q_{i,\nu} \ \xi_i(\rn) = 0$$

For the quadratic part, considering that the scatterers are identical
and the scattering events independent:
\begin{eqnarray*}
\langle \Delta \phi_{NA}^2 \rangle &=& \sum_{\nu=1}^{N_p} \sum_{\nu'=1}^{N_p} \left\langle
\left[ \qn \cdot \bm{\xi}(\rn) \right] \cdot
\left[ \bm{q}_{\nu'} \cdot \bm{\xi}(\bm{r}_{\nu'}) \right] \right\rangle\\
&=& \sum_{\nu=1}^{N_p} \left\langle
\left[ \bm{q}_{\nu} \cdot \bm{\xi} (\bm{r}_{\nu}) \right]^2
\right\rangle = N_p \left\langle
\left[ q \xi \cos \psi \right]^2 \right\rangle
\end{eqnarray*}
where $\psi$ is the angle between $\bm{q}$ and $\bm{\xi}$, which
verifies $\langle \cos^2 \psi \rangle = \frac{1}{3}$. Consequently,
\begin{equation*}
\langle \Delta \phi_{NA}^2 \rangle = \frac{1}{3}
    N_p \langle q^2\rangle \langle \xi^2 \rangle
\end{equation*}
We can furthermore compute $q^2$, using the scattering angle $\theta$
between $\bm{k}_{\nu - 1}$ and $\kn$ (see
Fig.~\ref{fig:def_r})~\cite{Weitz1993}:
\begin{equation*}
\langle q^2 \rangle = \left\langle \left[ 2 k \sin
  \frac{\theta}{2} \right]^2 \right\rangle = 4 k^2 \left\langle
\frac{1 - \cos \theta}{2} \right\rangle = 2 k^2 \frac{\ell}{\ell^*}
\end{equation*}
Finally, using the total path length $s=\sum_\nu l_{\nu}$ and the
relationship $N_p = s/\ell$ which holds in the multiple scattering
limit $s \gg \ell$, we obtain
\begin{equation}
\langle \Delta \phi_{NA}^2 \rangle = \frac{2}{3} k^2 \langle \xi^2
\rangle \frac{s}{\ell^*}
\end{equation}

%%%%%%%%%%%%%%
\subsubsection{Superposition of an affine field and a random motion}
If the displacement of the scatterers is the result of the
superimposition of an affine motion described by the displacement
field $\mathbf{u}(\rn)$ and a non-affine motion
$\bm{\xi}(\rn)$~\cite{Wu1990}:
\begin{equation}
\rn(t_2) = \rn(t_1) + \mathbf{u}(\rn) + \bm{\xi}(\rn).
\end{equation}

\begin{figure}[htbp]
\includegraphics[width=0.6 \columnwidth]{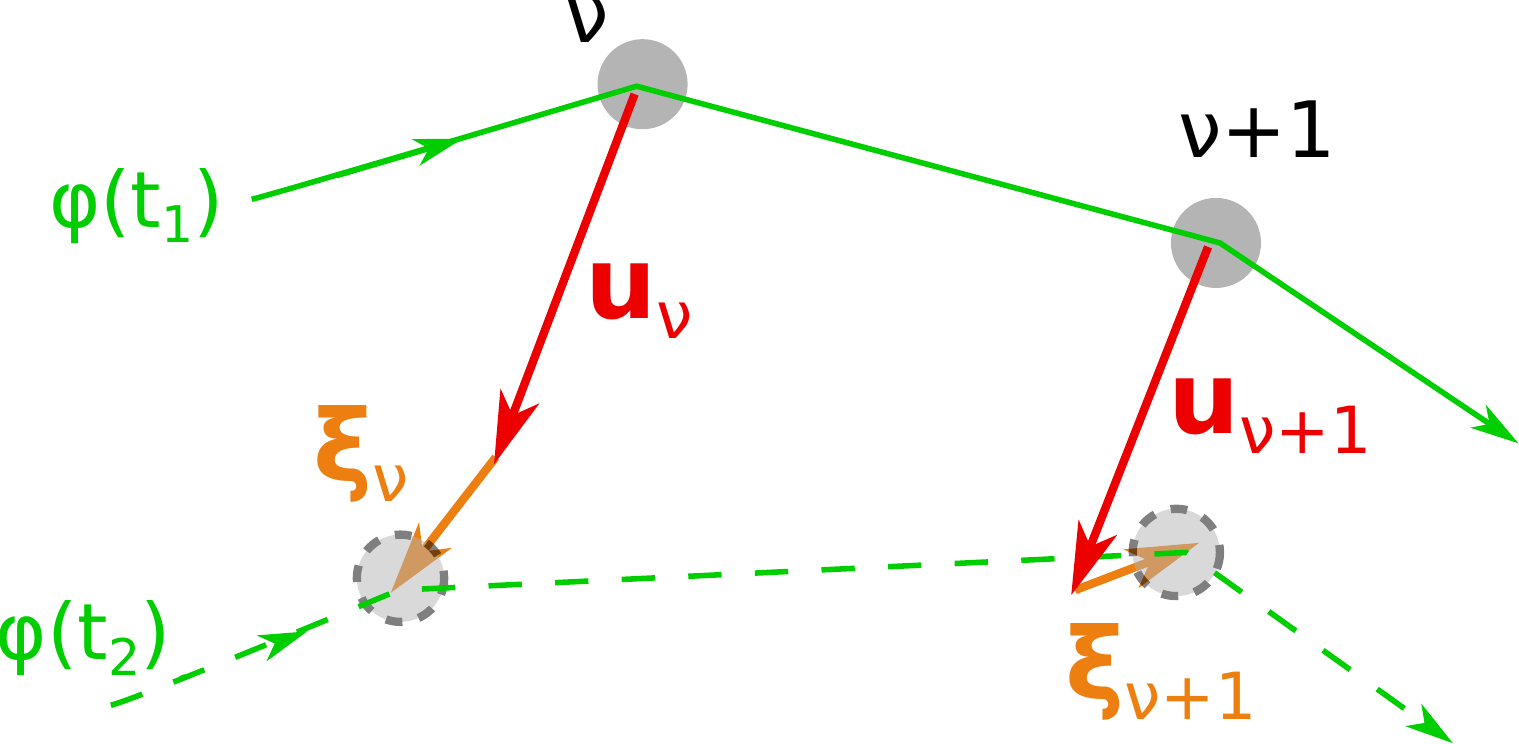}%
\caption{Schematic of the phase variation on a section of a path in
  the case of a superposition of a correlated motion given by the
  displacement field $\bm{u}$ and some uncorrelated motion of the
  scatterers given by $\bm{\xi}$.\label{fig:detail_phase_ANA}}%
\end{figure}

The total phase variation then is:
\begin{eqnarray*}
  \Delta \phi_T  &=& \sum_{\nu =0}^{N_p} \bm{k}_{\nu} \cdot \left[ \bm{u}(\bm{r}_{\nu + 1}) -
    \bm{u}(\bm{r}_{\nu})\right] + \sum_{\nu =1}^{N_p} \bm{q}_{\nu}
  \cdot \bm{\xi}(\mathbf{r}_{\nu})\\
\end{eqnarray*}

The total phase shift $\Delta \phi_T$ is thus the sum of a term
originating from the affine displacement $\Delta \phi_A$ and one due
to the non-affine motion $\Delta \phi_{NA}$:
$$\Delta \phi_T = \Delta \phi_A + \Delta \phi_{NA}$$

From the two previous parts, we can then deduce:
\begin{equation}
\langle \Delta \phi_T \rangle = \langle \Delta \phi_A \rangle
\end{equation}

As there is no correlations between the affine and the non-affine
motion, we have
\begin{equation}
\langle \Delta \phi_T^2 \rangle = \langle \Delta \phi_A^2 \rangle +
\langle \Delta \phi_{NA}^2 \rangle
\end{equation}

%%%%%%%%%%%%%%
\subsubsection{Expression of the correlation function}
Finally, in the case of the superposition of an affine displacement
due to the strain field given by the tensor $\bm{U}$ and a non-affine
motion characterized by the mean square displacement $\langle
\xi^2\rangle$, we obtain from Eq.~\eqref{eq:exp_Dphi}:
\begin{eqnarray}
\langle \exp \left(j \Delta \phi_T \right) \rangle &=& \exp \left( j
\frac{ks}{3} \text{Tr}\left( \bm{U} \right) \right) \times\\ & & \exp \left(
-k^2 s \ell^* \left[ f\left( \bm{U} \right) + \frac{1}{3}
  \frac{\langle \xi^2\rangle}{(\ell^*)^2}\right] \right) \nonumber
\end{eqnarray}
with $f\left( \bm{U} \right) = \frac{\beta - \chi}{2l^*} \left( \sum_i
U_{ii}\right)^2 + \frac{\beta}{l^*} \sum_{i,j} U_{ij}^2$.

We will see in the next section that there is a particular interest
 in the backscattering configuration. In this geometry,
the integral of Eq.~\eqref{eq_int_gE} weighted by the length
distribution of the paths can be computed. In the case of uncorrelated
motion of the scatterers, $g_E$ is given in good approximation
by~\cite{Pine1988,Weitz1993,Pine1990}:
\begin{equation}
\vert g_E(1,2) \vert \approx \exp \left(-\eta k \sqrt{\langle
    \xi^2\rangle} \right)\label{eq:G_BS}
\end{equation} where $\eta$ is a numerical factor of
  order 2 taking into account boundary conditions and polarization
  effects~\cite{Mackintosh1989}. We can extend this solution to the
  case of the superposition of an affine and a non-affine motion and
  deduce~\cite{Erpelding2008}:
\begin{equation}
\vert g_E(1,2) \vert \approx \exp(-\eta k l^* \sqrt{3f(\mathbf{U}) +
  \langle \xi^2\rangle / (\ell^*)^2}).
\end{equation}

Using the Siegert relation~\eqref{eq:siegert} we thus obtain
$$g_I - 1 \propto \exp(- 2\eta k l^* \sqrt{3f(\mathbf{U}) + \langle
  \xi^2\rangle / (\ell^*)^2}).$$ Experimentally it is convenient to
compute a normalized correlation function:
\begin{equation}
G_I (1,2) =\frac{\left\langle I_1 I_2 \right\rangle - \left\langle I_1
  \right\rangle \left\langle I_2 \right\rangle}{\sqrt{\left\langle
    I_1^2 \right\rangle - \left\langle I_1 \right\rangle^2}
  \sqrt{\left\langle I_2^2 \right\rangle - \left\langle I_2
    \right\rangle^2},} \label{eq:def_gI_exp}
\end{equation}
$G_I$ is proportional to $g_I - 1$ so that we await a dependence:
\begin{equation}
G_I(1,2) \approx \exp(-c (\bar{\epsilon} + \bar{\xi})) \label{eq_gI}
\end{equation}
where $\bar{\epsilon}$ is a scalar representative of the amount of
deformation in the material linked to the quadratic invariants of the
strain tensor and $\bar{\xi}$ corresponds to the amount of
uncorrelated motions in the material. The order of magnitude of the
constant $c$ can be estimated depending on the knowledge of the
scattering process in the material. For Mie scatterers, using $\eta=2$
we expect: $$c \bar{\epsilon} = \frac{8 \pi}{\lambda} \ell^*
\sqrt{\frac{2}{5}} \sqrt{\frac{1}{2} \left( \sum_i U_{ii}\right)^2 +
  \sum_{i,j} U_{ij}^2}$$ Practically, we use $c=8\pi\sqrt{2/5}
l^*/\lambda$ to estimate the amount of deformation from the normalized
intensity correlation~\cite{Erpelding2013,Amon2012}.

To discriminate between the elastic part $\bar{\epsilon}$ and the
plastic one $\bar{\xi}$, oscillatory loading is usually used allowing
to identify a reversible part and an irreversible
one~\cite{Erpelding2010a}. A very successful use of such cyclic
loading combined with DWS consists in detecting \emph{echoes} in the
correlation function while applying oscillatory shear strain to
emulsions~\cite{Hebraud1997}, foams~\cite{Hohler1997} or colloidal
glasses~\cite{Petekidis2002}. The loss of magnitude of those echoes
allows to measure the plastic part of the deformation at each
cycles~\cite{Hebraud1997,Hohler1997,Petekidis2002}. A more indirect
way to separate different contributions consists in tuning the
wavelength of the light probe to compensate homothetic displacements
of the scatterers~\cite{Crassous2009}. It is then possible to separate
the affine deformation from the non-affine
deformation~\cite{Crassous2009}.

%%%%%%%%%%%%%%%%%%%%%%%%%%%%%%%%%%%%%%%%%%%%%%%%%%%%%%%%%%%%%%%%%%%%%%%%%%%%%%%%%%%%%%%%%%%%%%%%%
\subsection{Spatial resolution} \label{spatial}

\subsubsection{Principle}
It could seem surprising and even impossible that imaging can be
performed when multiple scattering is at play. The principle is the
following one. First and most importantly, we exploit the fact that in
the backscattering configuration most of the paths are short: half the
photons exit the sample at a distance $\leq 2.7 \ell^*$ from their
entering point~\cite{Baravian2005,Erpelding2008,Zakharov2010}. This
feature could be considered as a drawback as the multiple scattering
assumption underlying the theory might fail down. Practically, the
exponential dependence of eq.~\eqref{eq:G_BS} corresponds to the
observations, the factor $\eta$ taking into account phenomenologically
of the fact that the first steps of the photons entering in the sample
before their full randomization can not be describe in the diffusive
model~\cite{Pine1988,Pine1990,Weitz1993}. The factor $\eta$ also
depends on the polarization of the detected
light~\cite{Maret1987,Mackintosh1989,Pine1990}. Indeed, by selecting
for the detected light a polarization parallel to (resp.
perpendicular to) the polarization of the incident beam, one can
select shorter (resp. longer) photon paths. In those two
configurations, the exponential dependence still hold, the order of
magnitude of $\eta$ being the same, around 2. Note that in a typical
setup the backscattering angle between the incident beam and the
source is always much larger that the one needed to detect enhanced
backscattering. To conclude, in the backscattering geometry, we expect
to be able to collect information mostly from small volumes of typical
size fixed by $\ell^*$, the practical volume scanned being modified
when selecting the polarization. We have tested this imaging technique
on several elastic scattering materials under known loading conditions
showing the accuracy of the method~\cite{Erpelding2008,Erpelding2013}.

Second, we use a near field speckles set-up which allows to image the
side of a sample. Historically, study of the speckle fluctuations in
the image plane has been extensively studied in biomedical
context~\cite{Brier2001}, but generally the dynamics is too fast to be
resolved~\cite{Zakharov2010}: the method can be directly applied only
in the case when speckle dynamics vary slowly between the two
states~\cite{Erpelding2008,Zakharov2010,Sessoms2010a}.

Finally, we use a multispeckle scheme to compute correlations between
two states of deformation of a material~\cite{Viasnoff2002} (see
Sec.~\ref{sec:DWS_princ}). Because of the use of CCD cameras in order
to collect simultaneously numerous speckle images, dark noise
corrections are necessary when computing the correlation
functions. The procedure to perform such corrections is described in
References~\cite{Cipelletti1999,Djaoui2005}.

\begin{figure}[htbp]
\includegraphics[width=1.\columnwidth]{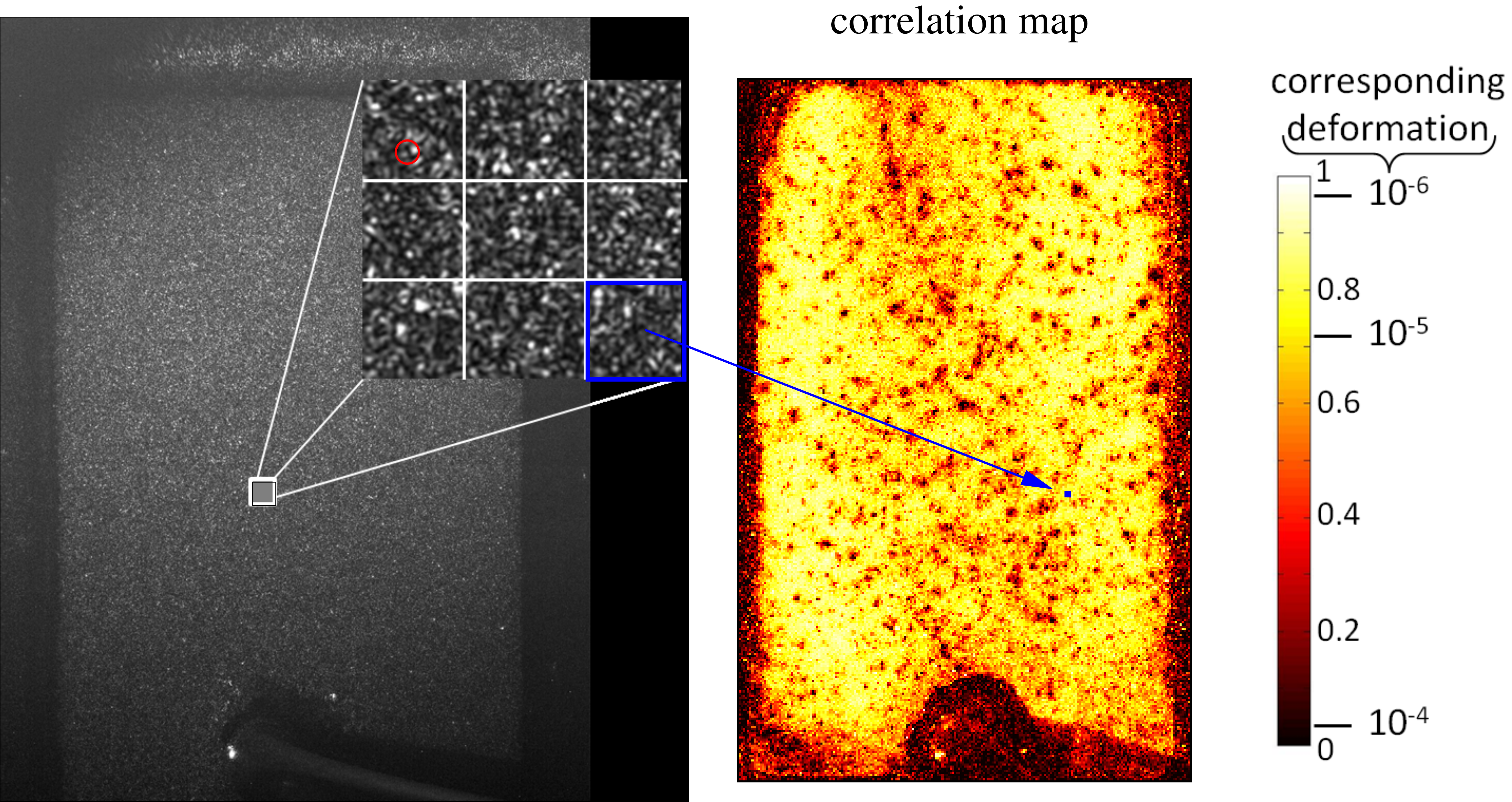}%
\caption{Left: raw experimental speckle image from an assembly of
  glass beads of diameter $d=90~\mu$m (in red in the inset). Right:
  correlation map computed using Eq.~\eqref{eq:def_gI_exp}. Each
  pixel, as the one colored in blue, is obtained from a multispeckle
  average on small zones from two raw images. The order of magnitude
  of the deformation indicated on the colorscale is obtained from
  Eq.~\eqref{eq_gI}.\label{fig:princ_DWS}}%
\end{figure}

The principle of the method is as following: the sample is illuminated
by an expanded coherent beam and a side of the sample is imaged using
a camera at different states of the material. The images are then
divided in areas called metapixels corresponding to a size $\ell^*
\times \ell^*$ on the sample (see Fig.~\ref{fig:princ_DWS}). The
aperture of the diaphragm is tuned to have enough speckles in each
area for the ensemble averages (see Sec.~\ref{sec:tutorial} for a
tutorial on the setup and the adjustments). The correlation
function~\eqref{eq:def_gI_exp} is computed for each metapixel between
two images, \emph{i.e.} two states of deformation of the
material. Each correlation function $G_I(1,2)$ of the final map
corresponds to a deformation in a volume $(\ell^*)^3$ in the vicinity
of the imaged side of the sample. As the light rays explore the bulk
of the material on thickness of few bead diameters, the method
described here differs from techniques based on speckles arising from
mere surface irregularities which provide only information on the
surface dynamics~\cite{Dainty1984}.

An example of a raw experimental picture and of a correlation map from
the experiment described in Sec.~\ref{sec_biaxial} are shown on
Fig.~\ref{fig:princ_DWS}. On the raw image, the glass beads are not
visible (a red circle in the inset indicates the size of the glass
beads in this experiment), the granular pattern of the image being
only due to the speckles. A correlation map obtained from computation
of correlations between two successive images is shown on right of
Fig.~\ref{fig:princ_DWS}. The colorscale for the correlation map is
given on the right: light color (white or light yellow) corresponds to
a correlation close to 1, \emph{i.e.}  deformation smaller than
$10^{-6}$; dark color (black) corresponds to a correlation close to 0,
\emph{i.e.} deformation larger than $10^{-4}$. The values of the
deformation are obtained using $\frac{- \ln G_I}{c}$,
i.e. Eq.~\eqref{eq_gI}.

\subsubsection{Dimensioning of the setup}
The size of the speckles $l_c$ can be chosen independently of the
magnification chosen to image the sample. The optimal size of the
speckles is the result of a balance between the fact that the
coherence areas have to be larger than the size of a pixel of the
camera and that for too large speckles, the information provided by
different pixels of the camera is redundant. Viasnoff et
al.~\cite{Viasnoff2002} have shown that the optimal speckle spot
diameter $l_c$ is of 3 pixels. Practically, we chose coherence areas
of sizes between 2 and 3 pixels (see Sec.~\ref{sec:tutorial}).

The optimal spatial resolution is obtained when a meta-pixel in the
image corresponds to a size $\ell^*$ on the object. The choice of the
lens magnification is then the result of a compromise between the
number of speckles in a metapixel $\gamma_t \ell^* / l_c$, which
determines the statistics for ensemble averages and the size of the
area to be studied which is determined by the lens magnification
$\gamma_t$.

A detailed discussion of an example of dimensioning of a setup is
given in Reference~\cite{Erpelding2008}.

%%%%%%%%%%%%%%%%%%%%%%%%%%%%%%%%%%%%%%%%%%%%%%%%%%%%%%%%%%%%%%%%%%%%%%%%%%%%%%%%%%%%%%%%%%%%%%%%%%%%%%%%%%%%%%%%%%%%%%%%%%%%%%%%%%%%%%%%%%%%%%%%%%%%%%%%%%%%%%%%%%%%%%%%%%%%%%%%%%%%%%%%%%%%%%%%%%%%%%
\section{A tutorial experiment: thermal deformation of a granular
  material} \label{sec:tutorial}

Despite the apparent simplicity of the DWS experiment, certain
difficulties can arise when implementing it for the first time. In
this section we give detail description on how to set the experiment
in order to obtain reliable results. For the demonstration we
performed a spatially resolved DWS experiment on granular sample that
was locally subjected to dilational expansion under heating. We can
expect that the speckle evolves, inducing a decorrelation of the
scattered intensity.

\begin{figure}[htbp]
\begin{center}
\includegraphics[width=.8\columnwidth]{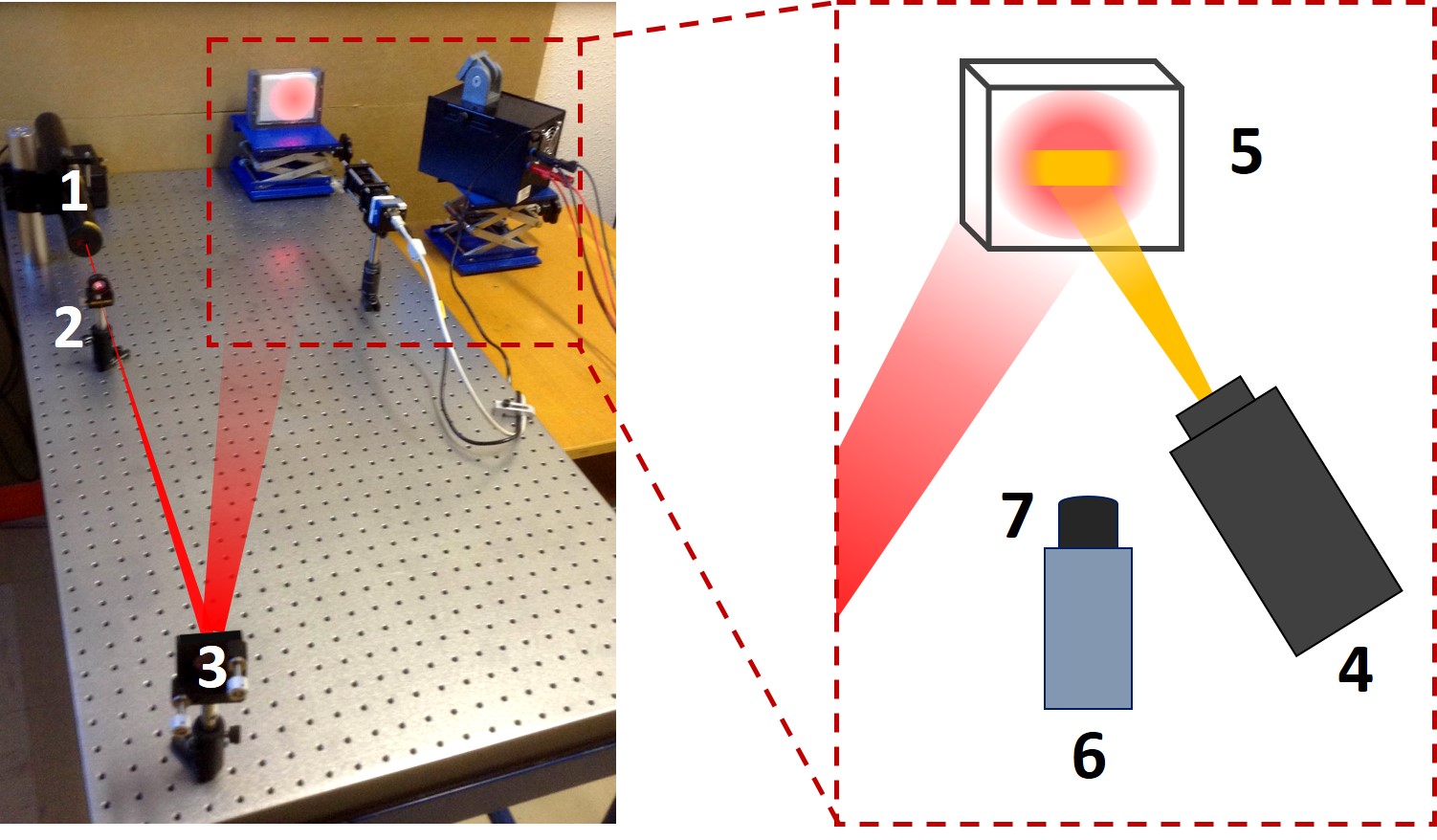}%
\caption{(a) Experimental set-up: 1-laser, 2-diverging lens, 3-mirror;
  (b) Zoom of the framed zone: 4-white source (heat source), 5-sample,
  6-CCD camera, 7-home made objective consisting in a lens, a
  diaphragm and an interferometric filter.}%
\label{fig:photo_setup}
\end{center}
\end{figure}

The detailed scheme of the set-up is shown in
Fig.~\ref{fig:photo_setup}. The heating source is a 75~W power halogen
lamp which is focused on the surface of a slab cell filled with glass
beads of mean diameter 90$\pm20$ $\mu$m. The high intensity of the
heating source allows an increase of the temperature in the vicinity
of the cell wall of 3-5 $^{\circ}$C in tens of seconds. The
deformation is imaged as explained in Sec.~\ref{spatial}. The plane
side of the sample is illuminated with a laser beam (Melles Griot
25-LHP-151-230, $\lambda=633$~nm, $P=5$~mW) that is preliminary make
diverged using a lens and a mirror. The area of the laser spot should
be large enough to cover all the area of interest in the
experiment. The image of the surface of the sample is formed
(magnification ratio $\gamma_t=0.2$) on the camera sensor (ProSilica
GC2450, resolution $2448 \times 2050$, square pixels of size $l_p=
3.45~\mu m$). A monochromatic filter is placed in front of the camera
to eliminate stray light.

\begin{figure}[htbp]
\begin{center}
\includegraphics[width=.75\columnwidth]{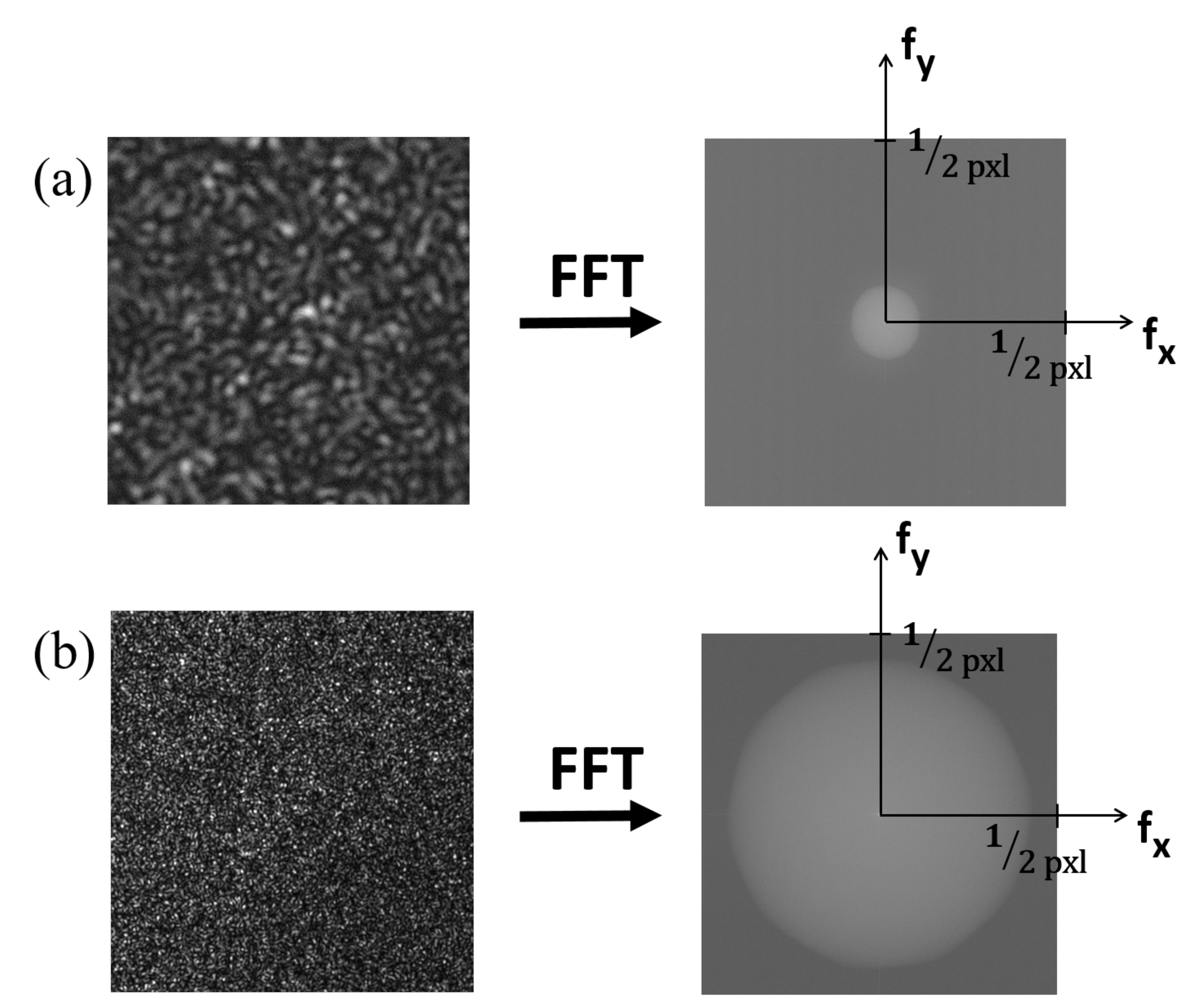}%
\caption{Image of a speckle pattern and its representation in spatial
  frequency domain: (a) image taken with a narrow aperture, (b) image
  taken with a wide aperture.}%
\label{fig:speckles}
\end{center}
\end{figure}

Concerning the camera setting, special attention should be paid to the
resulting size of a speckle spot, which can be tuned by adjusting the
aperture of a diaphragm in front of the camera. The optimal balance
between the signal-to-noise level and the optical contrast of the
image can be obtained for a speckle of $\simeq 2$ pixels in
diameter\cite{Viasnoff2002}. Its size can be measured by performing
2D-Fast Fourier transform (FFT) of a speckle pattern image. The
analysis gives the information on the frequency of the intensity
spatial distribution. The lower the spatial frequency the higher
number of pixels corresponds to one speckle and the more the image is
blurred. In contrast, high spatial frequency corresponds to a small
speckle size. This is illustrated in Fig.~\ref{fig:speckles}. The
Fourier transform of the scattered intensity is the exit pupil of the
imaging setup, i.e. the circular diaphragm~\cite{Goodman2007}. The
speckle size is $10.6$ and $2.3$ pixels for the images of
Figure~\ref{fig:speckles}(a) and (b) respectively. We set it to
$\simeq 2$~pixels in the following. Anomalous bright spots can be
observed in the speckle image due to specular reflections. Those spots
can be removed by slightly defocusing the imaging system and using a
polarizer in front of the camera crossed with the polarization of the
incident beam.

Another important parameter to be set is the frame rate. On the one
hand it should provide a good resolution time for the experiment and
fit to the rate of the dynamical process under investigation. On the
other hand, if images are acquired at too high a framerate, successive
speckle images are identical. Since we expect that thermal diffusion
occurs on time scale of few seconds, we set the frame rate to 1 image
per second.

\begin{figure}[htbp]
\begin{center}
\includegraphics[width=1.\columnwidth]{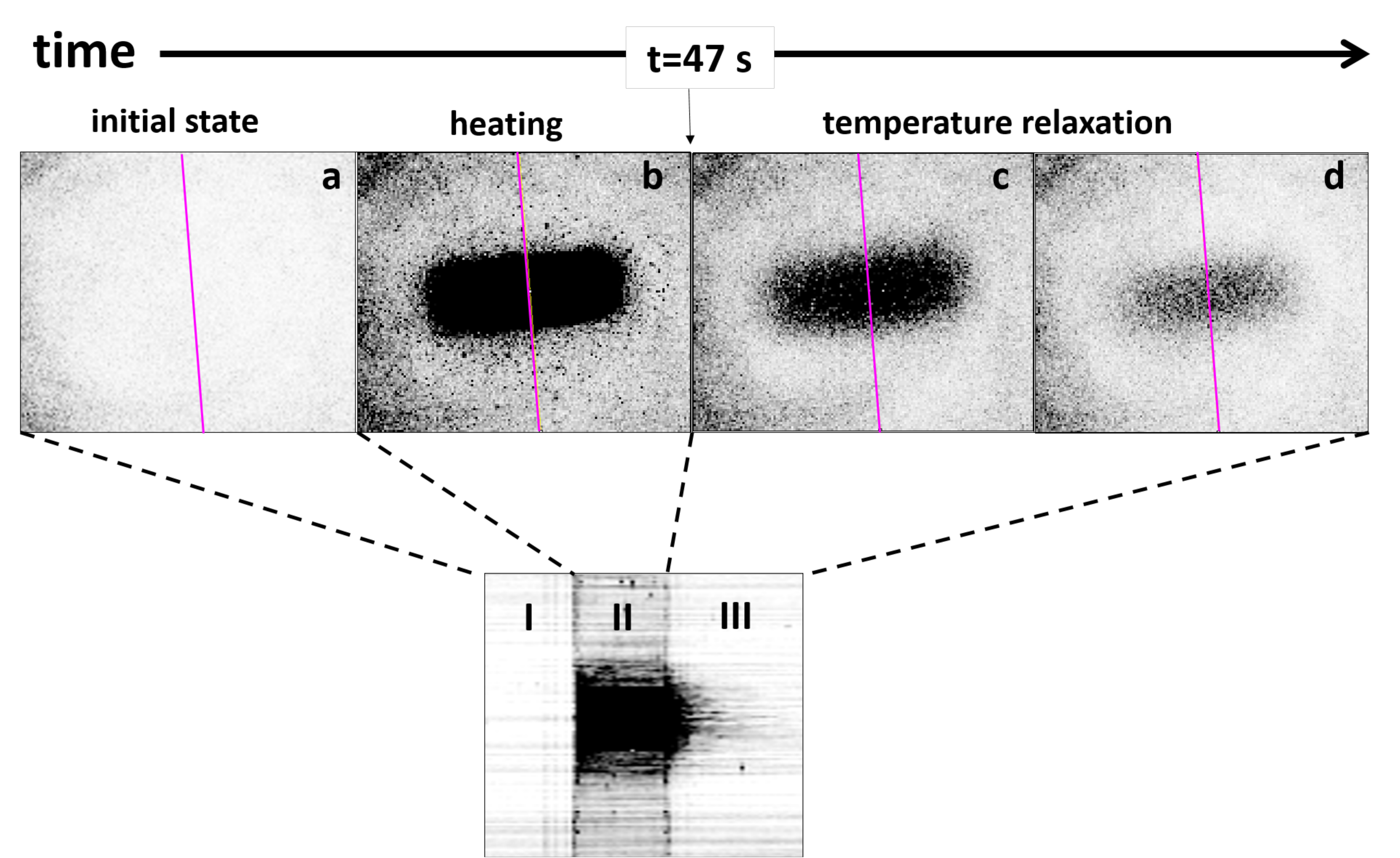}%
\caption{Upper line: Correlation maps corresponding to the heating
  experiment. (a) The white source is off and the correlation is
  maximal on the studied area. (b) Heating phase. The area where the
  white source is focused is fully decorrelated. At t=47s the white
  source is switch off again so that the relaxation process can be
  observed. (c) and (d) Correlations maps obtained during the
  relaxation. Bottom image: spatio-temporal graph obtained by stacking
profiles obtained on each maps along the magenta
line shown for the full recording.}%
\label{fig:stack}
\end{center}
\end{figure}

Images are taken before, during and after the sample
heating. Therefore the undisturbed state of the beads, loss of the
correlation and its following recovery have been tracked during the
experiment. We recall that to obtain the correlation map, the images
are first divided in metapixels and the correlation function are
computed for each metapixel between two different states of the
sample, here between two successive images.  Maps of correlation
function are done on zones of $16 \times 16$ pixels, corresponding to
$\simeq 50$ speckles spots. An order of magnitude of the noise on
$g_I$ depending on the size of the metapixel can be estimated using a
uniform map obtained by two successive images of the undeformed
sample. The ratio of the standard deviation and the mean value of
$g_I$ for different choices of metapixel size is shown in
Table~\ref{noise_gI}.
\begin{table}[htbp] \centering
\begin{tabular}{|p{2cm}|p{2cm}|}
\hline
4 pxl & $9.0\%$\\
\hline
8 pxl & $2.2\%$\\
\hline
16 pxl & $1.0\%$\\
\hline
32 pxl & $0.75\%$\\
\hline
\end{tabular}
\caption{Relative error on the value of $g_I$ in function of the size
  of the metapixel.} \label{noise_gI}
\end{table}
A size of $16 \times 16$ pixels is a good compromise between the noise
level and the final resolution of the correlation map. A length of
$16$ pixels on the camera corresponds to $16 l_p/ \gamma_t=270~\mu m
\simeq \ell^*$.

The correlation maps corresponding to each phases of the experiment
are shown in the upper part of the Fig.~\ref{fig:stack}. They are
presented in grayscale where the brightness of a pixel stands for the
value of the correlation function. Map (a) has been obtained before
the thermal deformation has been induced. It reveals the absence of
any dynamics in the system and the value of the correlation function
is close to 1. For all the pictures taken during the heating (e.g. map
(b)), the signal coming from the illuminated zone is totally
randomized and within this area the value of the correlation function
drops to 0. The true magnitude of the decorrelation due to beads
thermal expansion is not so large, however it is hidden under the
emission of the white light. The proper values of the correlation in
the system can be retrieved from the maps once the illumination is off
(Fig.~\ref{fig:stack}(c)). Then, the heat is conducted into the bulk
of the sample balancing the temperature gradient along the sample
surface. The activity in the system slows down and the correlation is
getting recovered as it can be seen on map~(d). It continues further
and at some point the situation comes back to state~(a).

\begin{figure}[htbp]
\begin{center}
\includegraphics[width=1.\columnwidth]{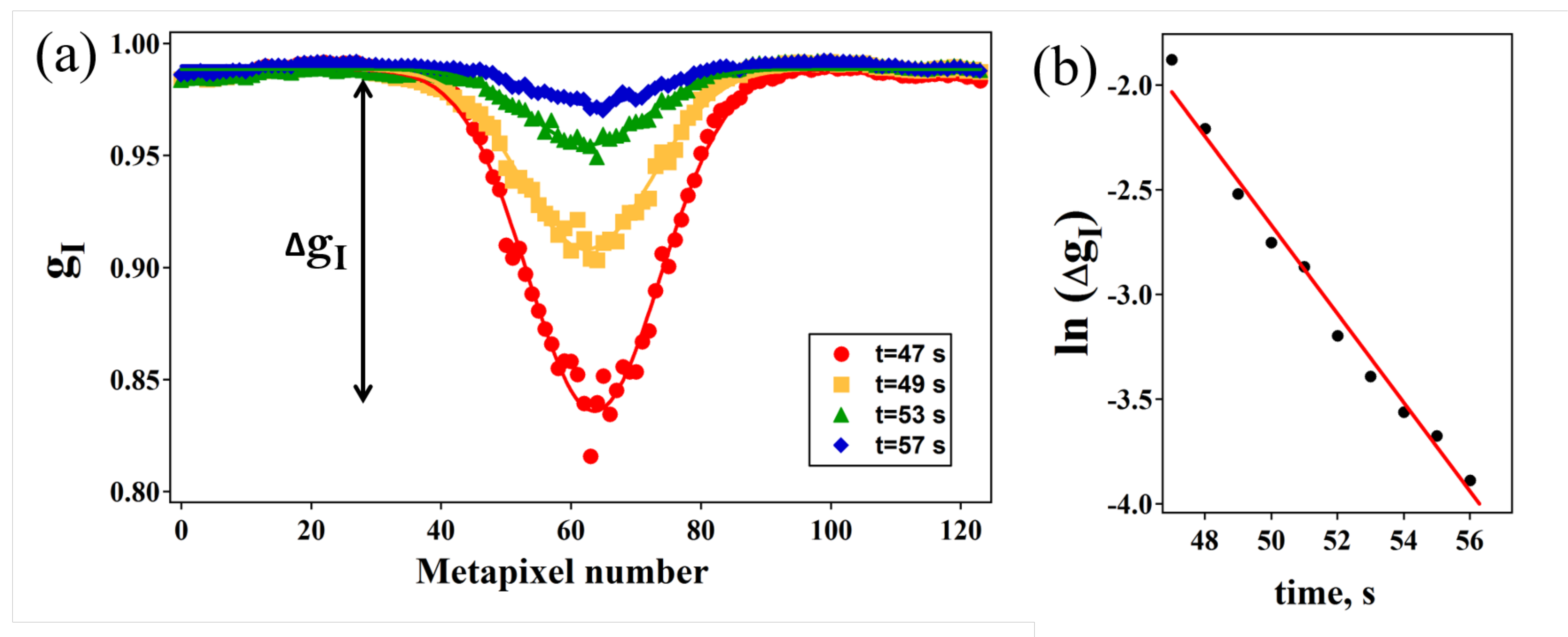}%
\caption{(a) Circles: profiles obtained from the correlation maps of
  Fig.~\ref{fig:stack} along the magenta line during the relaxation
  process. Different colors correspond to different times. Solid
  lines: fit of each profile by a Gaussian function. (b) Logarithm of
  the amplitude $\Delta g_I$ of the Gaussian fits as a function of
  time.}
\label{fig:exp_graph}
\end{center}
\end{figure}

Since the rate of the correlation recovery depends on the thermal
diffusivity of a medium, this parameter can be retrieved from the
obtained data. For that we monitor the variation of the correlation
function along a slice of the correlation map passing through the
heated zone. In Fig.~\ref{fig:stack} such a slice is indicated with
the magenta line. The temporal fluctuation of the correlation function
along this slice is presented in the bottom part of the figure by
stacking the lines of the full recording in a spatio-temporal
diagram. Here the number of pixel rows correspond to the length of the
slice and the number of columns corresponds to the number of images
taken during the experiment. The three phases of the experiment
(before heating, during and after) are clearly seen in the stack and
indicated as zones I, II and III, correspondingly. We are interested
in zone III, where the recovery of the correlation occurs. Several
profiles of the slices in zone III are plotted with circles in
Fig.~\ref{fig:exp_graph}(a). Different colors corresponds to
different images and therefore to different times. The profiles have
been fitted with the Gaussian function for which the amplitude of the
peak (noted as $\Delta g_I$) is associated with the magnitude of the
thermal deformation and consequently with the temperature
gradient. The amplitude $\Delta g_I$ decreases exponentially with time
as shown in Fig.~\ref{fig:exp_graph}(b). The relaxation time can be
measured: $\tau=4.5~s$. It corresponds to the time of heat diffusion
into the bulk. From reported values of thermal conductivity for glass
beads~\cite{Geminard2001} we may deduce a thermal diffusivity
$\nu=0.13~\text{mm}^2.s^{-1}$. We then find $\sqrt{\nu \tau}=0.7$~mm,
which is in agreement with the probed depth into the sample which is
few $\ell^*$.

Such experiment can be easily implemented with standard lab equipment
and allows to practice DWS measurement on a predictable
configuration. In the next section, we present some examples of
measurements that have been done on granular materials submitted to
different kind of loading.

%%%%%%%%%%%%%%%%%%%%%%%%%%%%%%%%%%%%%%%%%%%%%%%%%%%%%%%%%%%%%%%%%%%%%%%%%%%%%%%%%%%%%%%%%%%%%%%%%%%%%%%%%%%%%%%%%%%%%%%%%%%%%%%%%%%%%%%%%%%%%%%%%%%%%%%%%%%%%%%%%%%%%%%%%%%%%%%%%%%%%%%%%%%%%%%%%%%%%%
\section{Examples of applications}
Here, we present briefly measurements that have been done in different
configuration on granular samples. All the results presented in this
section have been extensively described and discussed in other
publications. We present them to illustrate the potential of the
method.

%%%%%%%%%%%%%%%%%%%%%%%%%%%%%%%%%%%%%%%%%%%%%%%%%%%%%%%%%%%%%%%%%%%%%%%%%%%%%%%%%%%%%%%%%%%%%%%%%
\subsection{Shear bands} \label{sec_biaxial}
\begin{figure}[hbtp]
\includegraphics[width=.75\columnwidth]{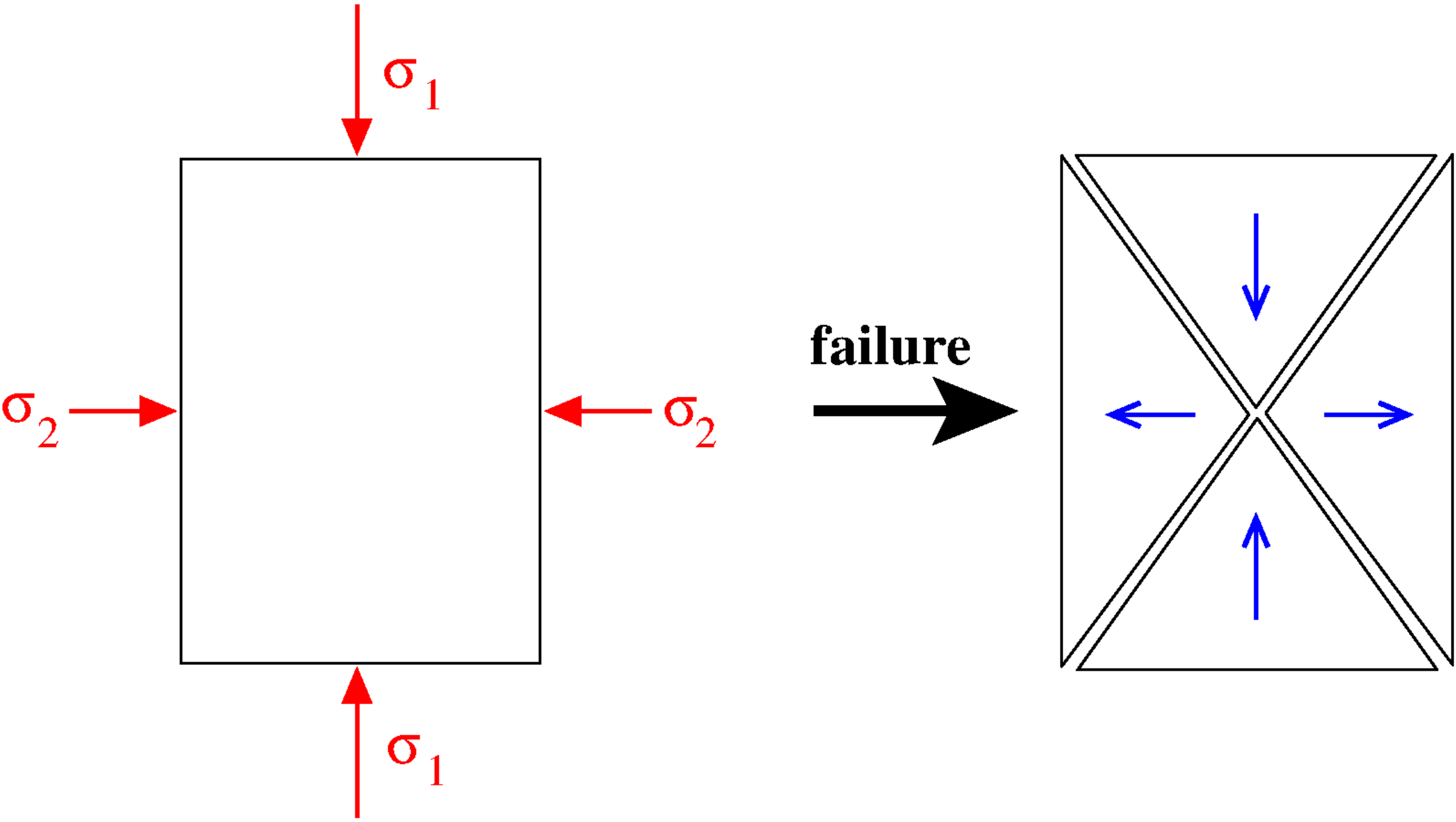}
\caption{Left: Principle of a biaxial test. Right: Schematic
  representation of the response of the material after failure.}
\label{fig:biaxial}
\end{figure}

\begin{figure}[hbtp]
\includegraphics[width=.8\columnwidth]{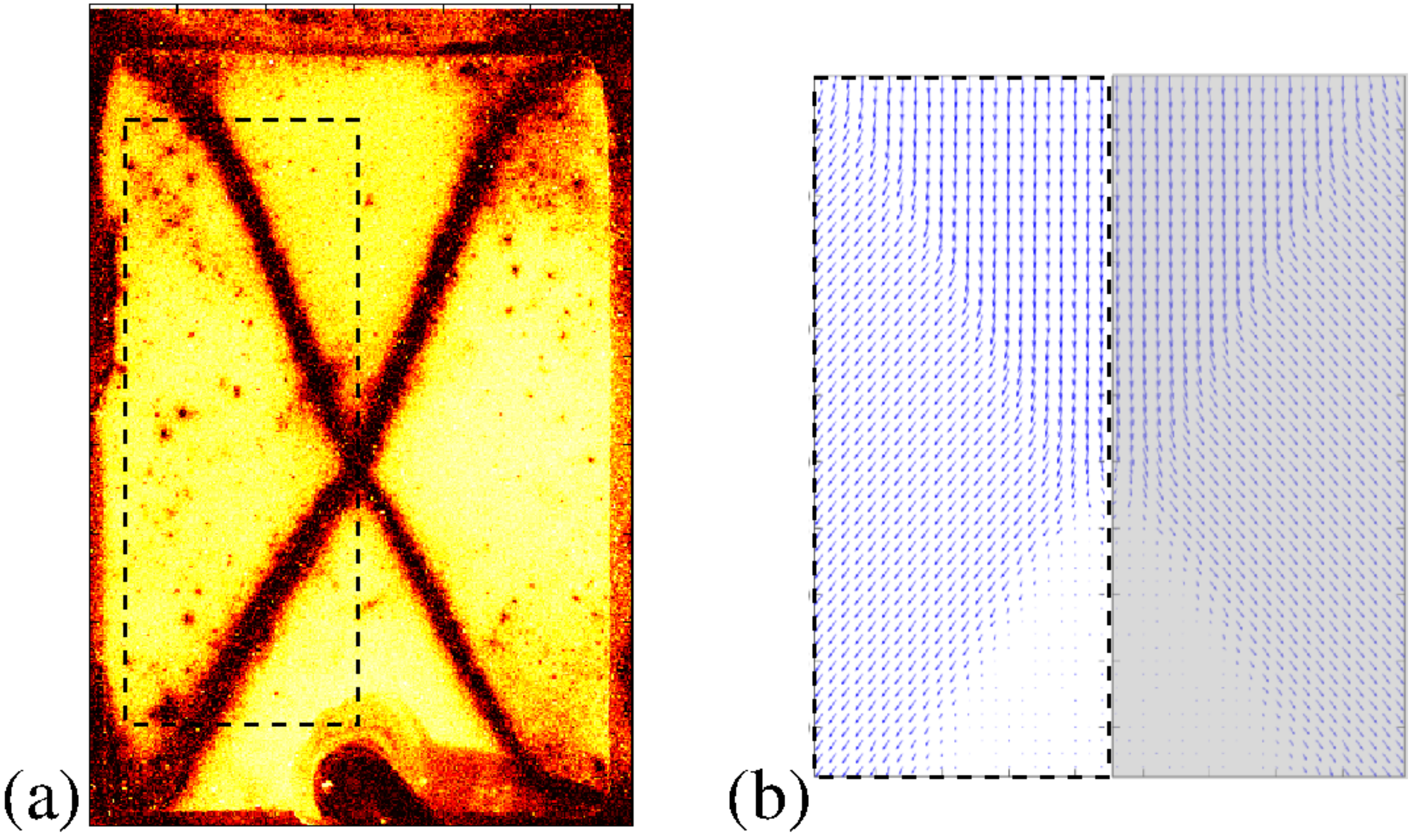}
\caption{(a) Correlation map obtained using DWS displaying two
  conjugated shear bands. (b) Averaged displacement field obtained
  using tracking. For practical reasons, only half of the sample has
  been imaged corresponding to the dashed area in (a). The shadowed
  part of the image has been added by symmetry for the sake of
  clarity.}
\label{fig:comp}
\end{figure}

A standard configuration to test failure of a granular material in
soil mechanics is the biaxial test~\cite{Desrues2004}: a sample
submitted to a confining pressure is uniaxially compressed. It is also
confined between two walls ensuring plane strain conditions (see
Fig.~\ref{fig:biaxial}(a)).
When the material fails it presents shear
bands. Fig.~\ref{fig:comp}(a) shows a typical correlation map obtained
using DWS after the failure of the material displaying two conjugated
shear bands. The mechanical response of the material can be
schematically considered as solid blocks moving relatively as shown in
the right part of Fig.~\ref{fig:biaxial}.

We have performed in this experiment two complementary measurements:
DWS measurements as described in the present review but also direct
tracking of particles. The second method allows a direct measurement
of the displacement field. An example of the average field is shown in
Fig.~\ref{fig:comp}(b). We observe that the displacement field
obtained by tracking coincides with the strain map.

The results allow to show that the DWS method gives a measurement of
\emph{relative} displacements, explaining the fact that the part of
the material that are in solid translation stay correlated in
average. Indeed, a solid translation of 1 $\mu$m of the scatterers
results in a translation of the speckle pattern in the image plane of
less than 3~\% of a pixel. Consequently, the contribution of this
translation to the decorrelation is negligible in this particular
study. Nevertheless, it is possible to identify the solid translation
of the speckle pattern complementary to its deformation in order to
perform Particle Image Velocimetry~\cite{Cipelletti2013}.

A detailed study of the response of material during this test can be
found in References~\cite{LeBouil2014a,LeBouil2014b,Nguyen2016}.

%%%%%%%%%%%%%%%%%%%%%%%%%%%%%%%%%%%%%%%%%%%%%%%%%%%%%%%%%%%%%%%%%%%%%%%%%%%%%%%%%%%%%%%%%%%%%%%%%
\subsection{Inclined plane: micro-ruptures}
\begin{figure}[hbtp]
\begin{center}
\includegraphics[width=1.\columnwidth]{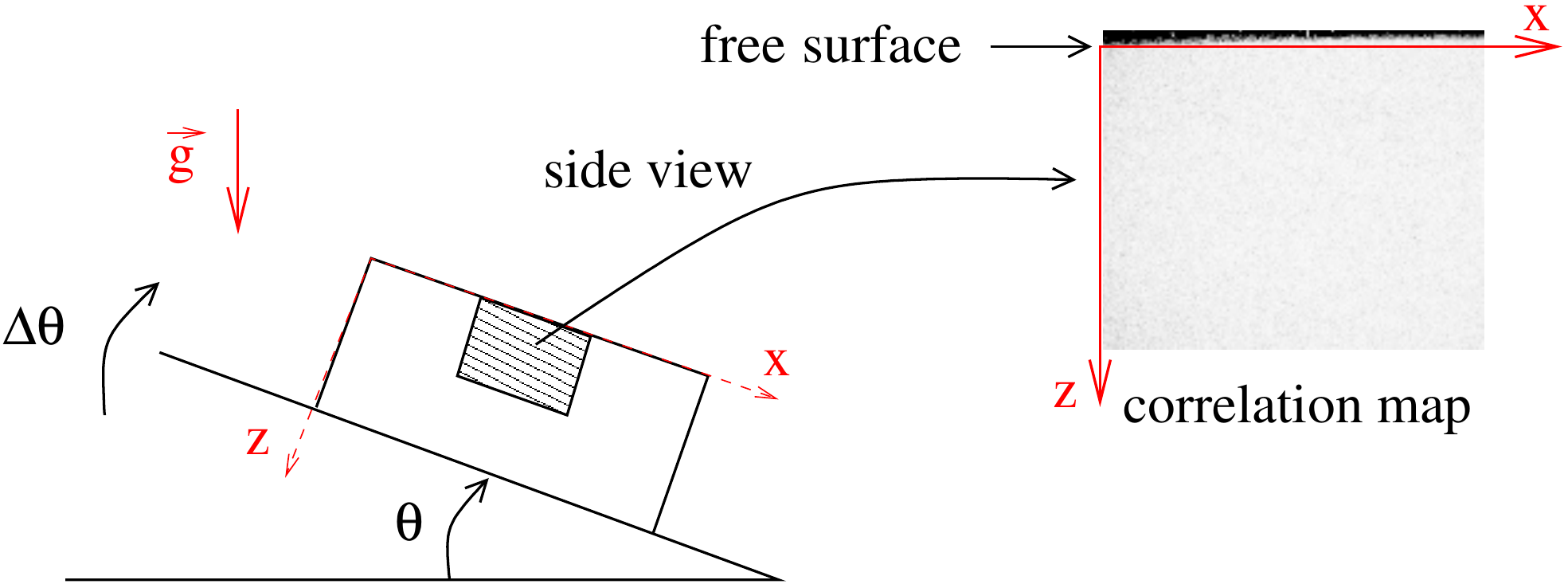}
\caption{Schematic of the experiment: a box filled with grains is
  quasi-statically inclined. Using DWS measurements we have scanned
  the side of the box so that the response with the depth in the
  sample can be studied. A correlation map at the beginning of the
  inclination process is shown.}
\label{prec_setup}
\end{center}
\end{figure}

Another standard configuration for studying failure in a granular
material is a progressive inclination a box filled with beads (see
Fig.~\ref{prec_setup}). The goal is then to identify the plastic
mechanisms that precede the destabilization of the pile in an
avalanche. In this system, small rearrangements as well as regular
large micro-ruptures have been evidenced before the
avalanche~\cite{Nerone2003,Kiesgen2009,Amon2013}.

\begin{figure}[hbtp]
\begin{center}
\includegraphics[width=1.\columnwidth]{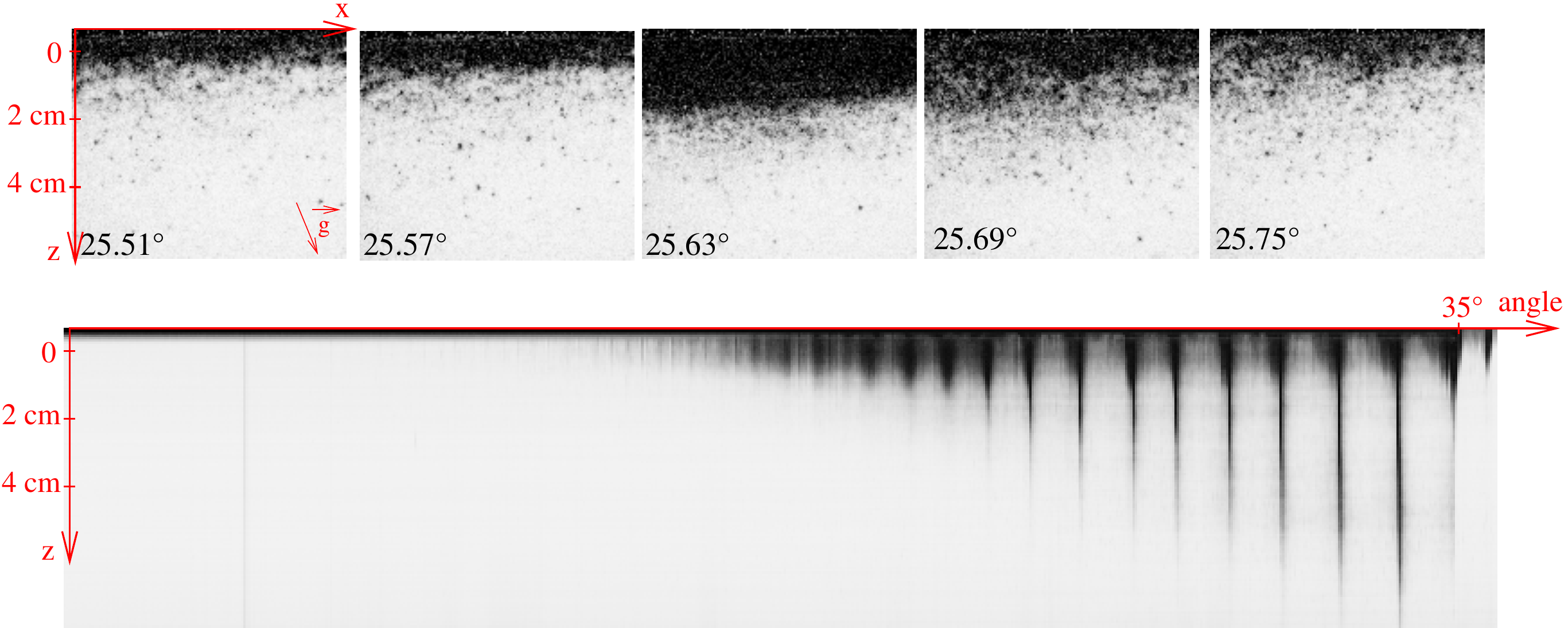}
\caption{Upper line: Successive correlation maps showing the formation
  of a micro-rupture. Bottom: spatio-temporal representation obtained
  by averaging the correlation maps at constant depth. The time axis
  is proportional to the inclination angle.}
\label{prec_film}
\end{center}
\end{figure}

Our observations using DWS evidenced that localized rearrangements are
present from the very beginning of the tilting process and occur at
all depths in the sample. At a given depth, the density of the
rearrangements increases with the shear, while at a given angle the
density of the rearrangements decreases with the depth. We have also
observed large events implying typically a part of the material
parallel to the surface. Such micro-rupture can be seen in
Fig.~\ref{prec_film} at 25.63$^\circ$. The successive correlation maps
surrounding the micro-rupture are shown. Those maps are incremental:
they are computed between two successive speckle images. It is then
possible to identify the details of the plastic processes at play
during such micro-ruptures.

The micro-rupture phenomenology is mainly a function of the depth $z$
under the free surface so that a spatio-temporal representation
obtained by averaging the values of the correlation at each depth $z$
allows a good visualization of the behavior of the sample during the
quasi-static tilting of the pile (see Fig.~\ref{prec_film}(b)). We
observe regularly spaced large events. The depth of those events
increases linearly with the angle of inclination until the
avalanche. Those events start from an angle typically around
15$^\circ$, independently from the type of material, revealing an
internal threshold well below the avalanche angle.

An extensive study in this configuration can be found in
Reference~\cite{Amon2013}. The unmatched resolution on deformation
allows here to identify processes of very small amplitude. When
studying precursors to catastrophic events, having access to such
minute local deformation is crucial to resolve the internal plasticity
before the failure.

%%%%%%%%%%%%%%%%%%%%%%%%%%%%%%%%%%%%%%%%%%%%%%%%%%%%%%%%%%%%%%%%%%%%%%%%%%%%%%%%%%%%%%%%%%%%%%%%%
\subsection{Heterogeneous deformation: response to a localized force}

\begin{figure}[hbtp]
\includegraphics[width=1.\columnwidth]{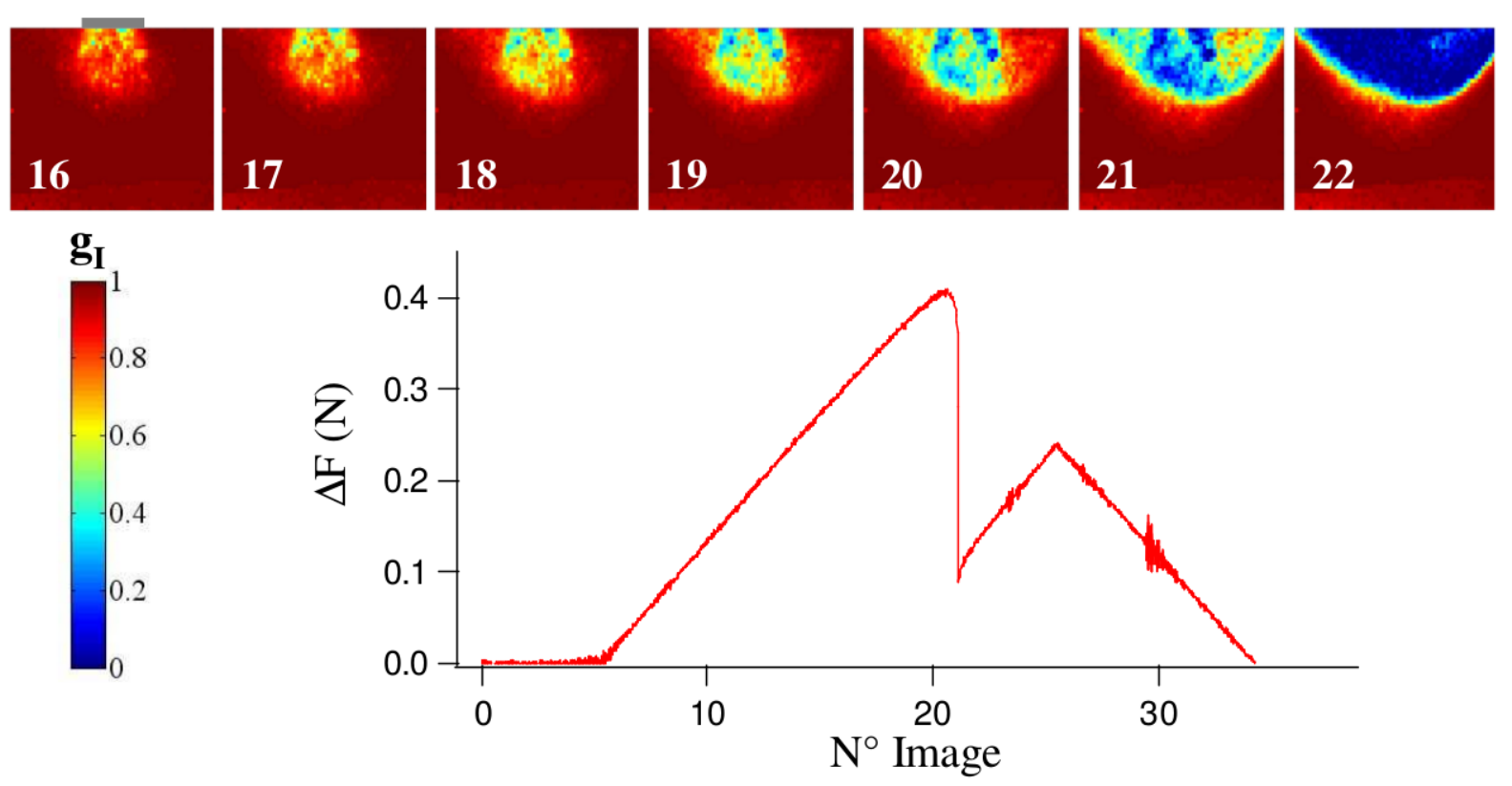}%
\caption{Correlation maps obtained during the force ramp shown at
  bottom right. The response evolves from an almost elastic one to an
  heterogeneous one. Finally failure occurs revealed in DWS as a large
  uncorrelated area.}%
\label{fig:marion_failure}%
\end{figure}

An unsolved question in granular matter is the problem of the elastic
limit of a non-cohesive granular material and of the response of this
material when submitted to cycles of force of a very small
amplitude. To separate the elastic, reversible, part from the plastic,
irreversible one in DWS measurements, an image of reference can be
used to compute all the correlations instead of studying only the
incremental deformation between consecutive images. It is then
possible to measure the recovered correlation during a cycle and to
separate it from the irreversible loss of correlation during a
cycle. To underline this difference in the present study in comparison
to the previous ones we use a different colormap here: red corresponds
to the maximal correlation ($g_I \approx 1$) while dark blue
corresponds to full decorrelation ($g_I \approx 0$) (see colorscale in
Fig.~\ref{fig:marion_failure}). In such study, a limitation can arise
from the temporal stability of the laser that can limit the maximal
time lag possible between two images. Typically, for non-stabilized
HeNe or Nd:YAG lasers~\cite{crassous2008}, the intensity correlation
function decreased of 10\% on a timescale of 100-1000~s. For studying
very slow dynamics, the use of stabilized laser allows to keep a
reference image for several hours.~\cite{Crassous2009}.

\begin{figure}[hbtp]
\begin{center}
\includegraphics[width=1.\columnwidth]{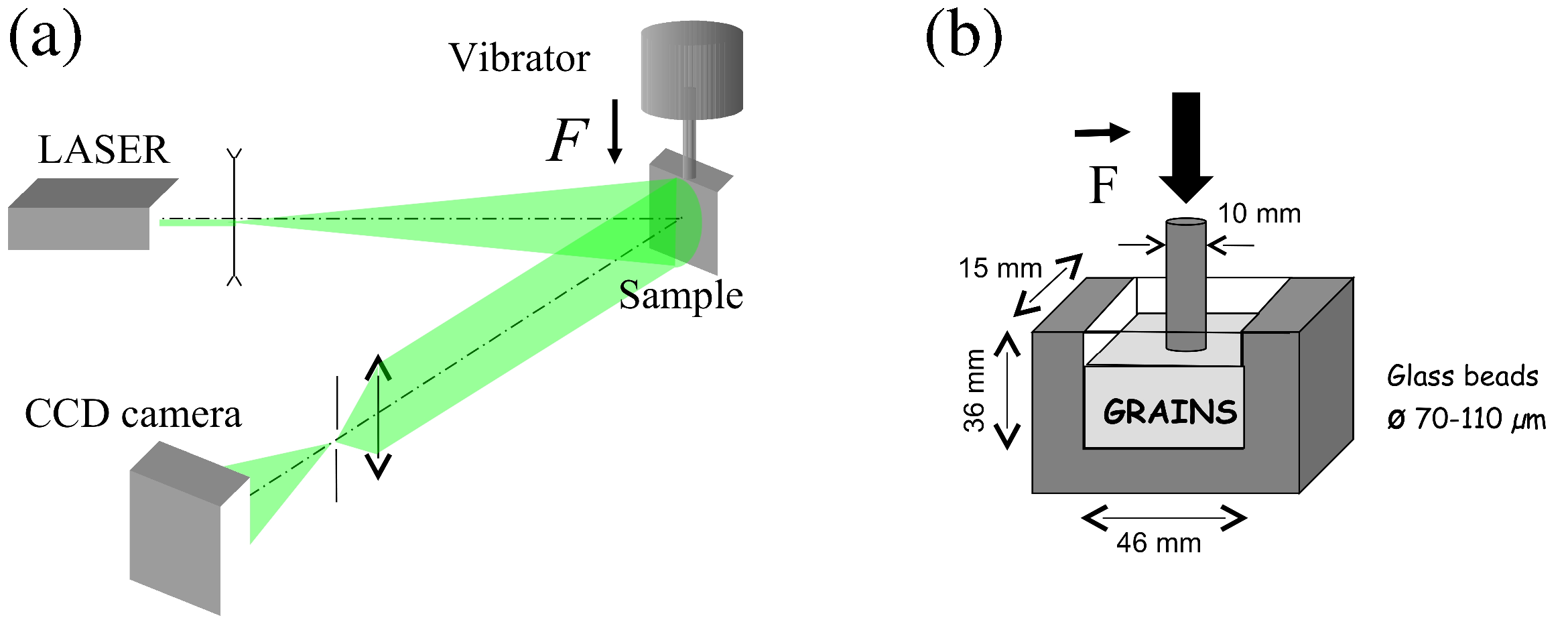}
\caption{From~\cite{Erpelding2010b} (a) Schematic of the setup. (b)
  Details of the sample.}
\label{marion_setup}
\end{center}
\end{figure}

We have studied the mechanical response of a granular pile to a
localized force and we have characterized experimentally the main
differences between the response of the granular sample and the one of
an elastic reference media~\cite{Erpelding2010a}. The experimental
setup is shown in Fig.~\ref{marion_setup}.

The upper row of Fig.~\ref{fig:marion_failure} shows the observed
spatial repartition of the deformation in the sample during a force
cycle for which we have observed a failure in the material. The
reference image is the initial speckle image corresponding to the
unloaded sample. We observe inhomogeneities in the otherwise regular
response that we interpret as precursors of the rupture. As in the
previous example, all the interest of DWS measurements is the
possibility to study the intermittent heterogeneous dynamics that
precedes the failure.

%%%%%%%%%%%%%%%%%%%%%%%%%%%%%%%%%%%%%%%%%%%%%%%%%%%%%%%%%%%%%%%%%%%%%%%%%%%%%%%%%%%%%%%%%%%%%%%%%%%%%%%%%%%%%%%%%%%%%%%%%%%%%%%%%%%%%
\section{Conclusion}
We have presented pedagogically how to measure deformation in granular
materials using DWS. We have given the basic tools both on the
theoretical side and on the very practical side for the reader to be
able to implement this method in the lab. We have given a short review
of the principle of Diffusing Wave Spectroscopy. We have discussed in
more details the case of deformation in granular materials with
special interest on how scattering take place in granular matter and
how to take into account uncorrelated motion in addition to an affine
deformation field when analysing the correlation functions. We have
presented in details a tutorial experiments with practical tips on the
adjustment of the setup. Finally we have briefly presented some
observations done using this method in various experimental setups
implying granular materials. This article should allows any beginner
in the field to implement the method rapidly.

\section*{Acknowledgements}
Part of the works presented here have been obtained during the PhD
Theses of Marion Erpelding and Antoine Le Bouil as well as during the
Master internship of Roman Bertoni. A. M. acknowledges postdoctoral
financial support from ESA.
%%%%%%%%%%%%%%%%%%%%%%%%%%%%%%%%%%%%%%%%%%%%%%%%%%%%%%%%%%%%%%%%%%%%%%
\bibliographystyle{unsrt}

\end{document}